# CHAPTER 7

# NICKEL OXIDE QUANTUM DOTS AND POLYMER NANOWIRES

Liudmila A. Pozhar

*PermaNature, Birmingham, AL 35242*

Home Address: 149 Essex Drive, Sterrett, AL 35147

Tel: (205) 678-0934

E-mail: lpozhar@yahoo.com; pozharla@yahoo.com

Running Title: **Nickel Oxide Nanostructures**

# SUMMARY

The virtual (*i.e.*, fundamental many body quantum theory-based, computational) synthesis method is used to establish electronic templates of about 30 non-stoichiometric nanosystems composed of nickel and oxygen atoms and ranging from about 6 Å to 6 nm in linear dimensions. Flexible and stretchable Ni-O bond in such structures accommodates various ratios of Ni to O atoms, and both antiferromagnetic and ferromagnetic spin alignments. Depending on synthesis conditions, smaller Ni-O quantum dots (QDs) composed of up to 14 atoms or so may have both types of spin alignments, while quantum-confined, quasi one dimensional Ni-O nanowires (QWs) appear to be nanopolymers with antiferromagnetic spin alignment. Ni-O bond flexibility and related ease of spin re-arrangement may facilitate physical mechanisms leading to the development or loss of exchange bias when such Ni-O quantum dots and wires (QDWs) interact with surfaces or each other at some thermochemical conditions. These structural and compositional flexibility is reflected by QDWs' molecular electrostatic potential (MEP) and electronic level structure (ELS). In particular, the direct optical transition energy (OTE) of the studied Ni-O systems may vary within an order of magnitude, and their electronic and magnetic properties can be finely tuned to match applications requirements by manipulations with synthesis conditions, the structure and composition of quantum confinement. The developed electronic templates are available upon request.

# 1. INTRODUCTION

Since 70's of the last century, nickel oxide nanoclusters and nanostructures enjoyed an impressive variety of applications. Thus, small $Ni_xO_y$ anions with $x$ ranging from 1 to 12 and $y$=0, 1, 2 produced by laser vaporization were thoroughly studied experimentally in conjunction with heterogeneous catalysis, formation of catalytic active sites and chemisorption (1, 2, 3). Novel colloidal composites achieved through growth of Ni/NiO core-shell nanoparticles were shown to incorporate multiple functionalities, including optical and magnetic manipulations, thermal response and molecular trapping (4). Photocatalytic hydrogen production and solar energy conversion systems were observed to be enhanced through the use of nickel oxide nanoclasters (5). However, synthesis of advanced nanomaterials for sensor, electronics and spintronics applications (6) - (16) is among the major fields greatly benefitted from exceptionally useful magnetic and electronic properties of nickel oxide nanoclusters and nanostructures, especially those fabricated in quantum confinement or on surfaces (17, 18). Recently, electron spin states of exchange-biased, core–shell Ni-O nanoclusters attracted significant interest as prospective physical carriers of information qubits (19), because such states can be precisely controlled. It has been understood that exchange bias phenomenon observed in Ni-O core-shell nanoclusters (20) may be used as a natural means of such control paramount for spintronics and electron spin-based quantum information processing.

Further advances in spintronics, quantum electronics, quantum computing and quantum information processing require a dramatic decrease in size of nanoclusters to only a few atoms. At the same time, recent experiments (19), (21) indicate that some useful effects, such as exchange bias, may be lost in such small, sub-critical size nanoclusters. Thus, first-principle understanding of structure - property relationships for and charge/spin localization in promising small Ni-O QDWs is extremely important both from fundamental and practical standpoints. This chapter is focused on electronic and magnetic properties of small Ni-O structures, in view of their utmost importance for nanoelectronics and emerging quantum information processing-based technologies.

Experimental assessment of synthesis conditions and properties of small Ni-O clusters is at the limits of contemporary experimental techniques (17, 18, 22). Moreover, understanding and interpretation of the available and emerging experimental results must be based on *ab initio* theoretical studies to reveal physical and chemical mechanisms behind observed properties of such systems. The existing first-principle theoretical studies of electronic and magnetic properties of small Ni-O clusters have used both density functional theory (DFT) and many body quantum theoretical methods [see, for example, Refs. (23) - (25)]. In available DFT-based theoretical studies Ni-O clusters are usually envisioned as elements of a bulk Ni-O lattice, and modeled either as surrounded by "point charges" (26), or as a periodic slab (2), to ensure that the electronic and magnetic properties of the clusters match those of the bulk Ni-O lattice. However, there is no fundamental justification behind expectations that the structure or/and properties of small systems composed of only few atoms should mimic those of the corresponding bulk systems. Moreover, even modern versions of DFT are non-variational theories, and thus cannot provide unambiguous computational results when applied to quantum mechanical systems [see a discussion and references in Ref. 27]. Thus, in this chapter electronic and magnetic properties of a large family of small Ni-O clusters are studied using first-principle, many body quantum theoretical methods and their computational realizations, and in particular the virtual synthesis method (27) - (31) discussed and used in previous chapters of this book. These methods do not require introduction of heuristic adjustments, such as exchange-correlation functionals in DT-based methods. The virtual synthesis approach exploits several approximations of increasing accuracy, including restricted and restricted open shell Hartree-Fock (RHF, ROHF and HF, respectively), configuration interaction (CI), complete active space (CAS) self-consistent field (SCF), and multiconfiguration SCF (MCSCF) approximations, as realized by GAMESS software package (32, 33) with the SBKJC standard basis set (34).

The "standard" optimization procedure of Refs. 32 and 33 has been applied to model nucleation of small Ni-O atomic clusters in the absence of any external fields exerted on the cluster atoms by foreign atoms or fields. Within the adopted approximations, this procedure converges to the equilibrium geometry and global minimum of the total energy of the optimized clusters in the absence of any external field and foreign atoms, that is, to the ground state of the corresponding molecules. Molecules synthesized in this fashion are called below vacuum molecules (*i.e.*, molecules synthesized virtually in physical "vacuum"). Recently, this standard optimization procedure has been modified (27) - (31), (35) to incorporate the major effects of quantum confinement and surfaces on molecular synthesis without detailed atomistic modeling of environment surrounding an optimized atomic cluster (see a detailed discussion in Chapter 1). In this approach, excluded volume effects, and to a significant degree, polarization effects are taken into account through spatial constraints applied to optimized systems. Such a modified procedure is necessary to bypass hardware and software restrictions that reduce applications of the many body quantum field theoretical methods to systems of about 100 many-electron atoms. In the modified optimization procedure, symmetry elements of bulk lattices and other structures are used to pre-design clusters' geometry and composition, and to keep the centers of mass of the cluster atoms constrained to specified positions in space. Subsequently, the standard total energy minimization is applied to such spatially constrained clusters to obtain their ground state energy and electronic structure of the constrained molecules (called below pre-designed molecules). This energy minimization process converges to a local minimum of the total energy of an optimized cluster that corresponds to spatial constraints applied to the cluster atoms, thus modeling effects of a confinement or surface on the synthesized molecule.

In calculations presented and discussed below, experimental values of 1.24 Å and 0.66 Å were used as initial covalent radii of Ni and O atoms. The virtual synthesis method was applied to configure a large number of non-stoichiometric, vacuum and pre-designed Ni-O molecules composed of 3 to 46 atoms. Understanding electronic and magnetic properties of such molecules is necessary to predict and control magnetism of advanced nickel oxide nanomaterials, and to predict and explain exchange bias effects at nanoscale.

## 2. MOLECULES DERIVED FROM $Ni_2O$ CLUSTER

The first two molecules (Table I) of this study are derived from an atomic cluster composed of one O and two Ni atoms. They are the linear vacuum MCSCF septet $Ni_2O$ (Fig. 1) and predesigned MCSCF triplet $Ni_2O$ (Fig. 2), both of which have a deep ground state energy minimum.

Being a septet, the vacuum molecule is synthesized at the limits of applicability of the used approximations, and may not be stable at experimental conditions. Both molecules are "ferromagnetic", with their uncompensated electron spins aligned in the same direction and localized on Ni atoms, and the electron charge accumulated in the vicinity of O atoms (Figs. 1c and 2c). The highest occupied molecular orbits (HOMOs) of these molecules are composed of $3d$ orbits of Ni atoms ($3d_{xy}$ and $3d_{xz}$, in the case of the triplet, Fig. 2b), with only minor contributions from $2p_x$ and $2p_z$ orbits of O atoms.

TABLE I.  Ground state energies and OTEs of small Ni-O molecules.

| No. | Molecule | RHF/ROHF ground state energy, Hartree | MCSCF ground state energy, Hartree | RHF/ROHF OTE, eV | MCSCF OTE, eV | RHF/ROHF spin multiplicity |
|---|---|---|---|---|---|---|
| 1. | Linear $Ni_2O$ | -352.635956 | -352.641378 | 3.5130 | 10.9036 | vacuum septet |
| 2. | Linear $Ni_2O$ | -352.430358 | -352.450052 | 1.2708 | 4.9838 | predesigned triplet |
| 3. | Triangular $Ni_2O$ | -352.406852 | -352.441908 | 1.9456 | 6.5443 | vacuum singlet |
| 4. | Triangular $Ni_2O$ | -352.380775 | -352.416856 | 6.1253 | 6.2695 | predesigned singlet |
| 5. | Square $Ni_2O_2$ | -368.161288 | -368.169562 | 0.0735 | 6.9607 | vacuum pentet |
| 6. | Square $Ni_2O_2$ | -368.107922 | -368.112991 | 8.7838 | 8.0164 | predesigned singlet |
| 7. | Modified $Ni_2O_2$ | -368.175174 | -368.216201 | 3.6354 | 2.7429 | vacuum septet |
| 8. | Modified $Ni_2O_2$ | -367.719865 | -368.067697 | 8.5797 | 2.3266 | predesigned singlet |
| 9. | Pyramidal $Ni_4O$ | -689.301239 | -689.385034 | 0.2177 | 0.1633 | vacuum triplet |
| 10. | Pyramidal $Ni_4O$ | -689.350295 | -689.378840 | 0.2176 | 0.8599 | predesigned pentet |
| 11. | Octahedral $Ni_4O_2$ | -705.020224 | -705.164661 | 0.4573 | 0.6504 | predesigned septet |
| 12. | Octahedral $Ni_4O_2$ | -704.865259 | -704.895966 | 1.1592 | 2.8273 | vacuum septet |
| 13. | Prismatic $Ni_6O_6$ | -1104.943626 | -1104.976724 | 3.9783 | 14.9390 | predesigned pentet |
| 14. | Prismatic $Ni_6O_6$ | -1104.806148 | -1105.045861 | 10.9635 | 14.3833 | vacuum pentet |
| 15. | QD $Ni_7O_6$ | -1272.671838 | -1272.528182 | 0.0381 | 3.0994 | predesigned pentet |
| 16. | QD $Ni_7O_6$ | -1273.167164 | -1273.256344 | 0.0952 | 0.4027 | vacuum triplet |
| 17. | QW $Ni_8O_6$ | -1440.716675 | -1440.911084 | 0.1360 | 0.0109 | predesigned triplet |
| 18. | QW $Ni_8O_6$ | -1441.561976 | -1441.654850 | 6.4110 | 1.0776 | vacuum singlet |

TABLE II.  RHF ground state energies and OTEs of small Ni-O nanopolymer quantum wires (QWs).

| No. | Molecule (RHF singlet) | RHF/ROHF ground state energy, Hartree | RHF/ROHF OTE, eV | Dipole moment (Debye) | QW length (nm) |
|---|---|---|---|---|---|
| 1. | $Ni_{10}O_8$ | -1808.2547340189 | 5.3418 | 0.011406 | 2.48 |
| 2. | $Ni_{12}O_{10}$ | -2176.1394863181 | 5.0450 | 2.408197 | 2.98 |
| 3. | $Ni_{14}O_{12}$ | -2544.0484383246 | 5.3362 | 0.006731 | 3.47 |
| 4. | $Ni_{16}O_{14}$ | -2911.9347598008 | 4.9089 | 3.392921 | 3.97 |
| 5. | $Ni_{18}O_{16}$ | -3279.8419509998 | 5.3389 | 0.014099 | 4.46 |
| 6. | $Ni_{20}O_{18}$ | -3647.7299132522 | 4.9334 | 4.005419 | 4.96 |
| 7. | $Ni_{20}O_{18}$ (triplet) | -3647.8951586947 | 0.7646 | 13.418265 | 4.96 |
| 8. | $Ni_{20}O_{18}$ (vacuum singlet) | -3647.9462432375 | 5.0069 | 4.399016 | 4.96 |
| 9. | $Ni_{24}O_{22}$ | -4383.5248289489 | 4.9634 | 4.530491 | 5.95 |

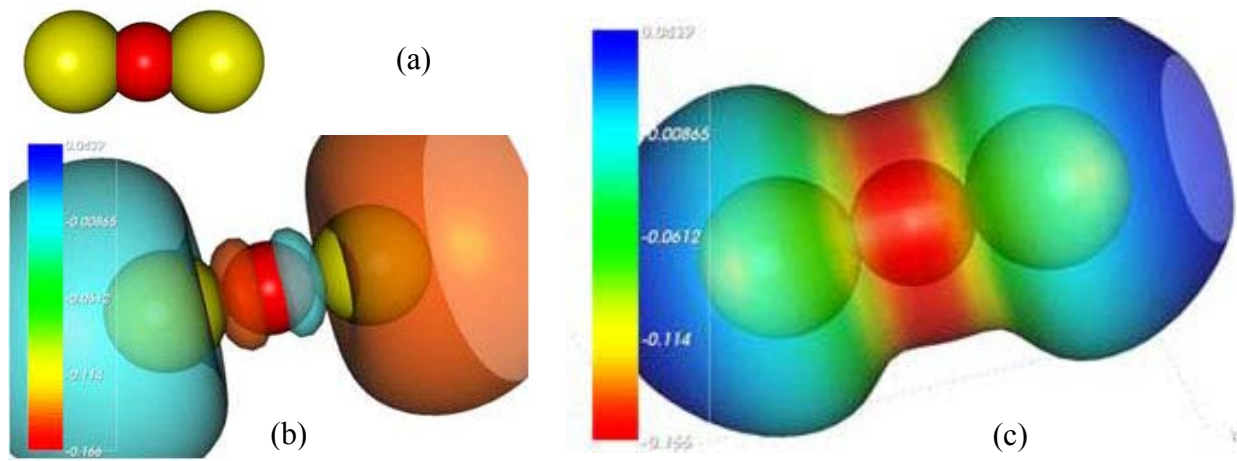

Fig. 1. (Color online) Linear vacuum MCSCF septet $Ni_2O$: (a) structure; (b) isosurface of the molecular electrostatic potential [MEP; values are varying from negative (red) to positive (deep blue)] corresponding to the fraction (cut) 0.01 of the maximum value (not shown) of the electron charge density distribution (CDD); (c) isosurface of the positive (red) and negative (blue) parts of the highest occupied molecular orbit (HOMO) corresponding to the isovalue 0.01. Ni and O atoms are represented by golden and red spheres, respectively. Electron charge (c) is accumulated in the vicinity of the oxygen atom (orange to yellow and to green regions). Atomic dimensions in (a) are to scale, and in (b) and (c) reduced. Other dimensions in the figures are to scale.

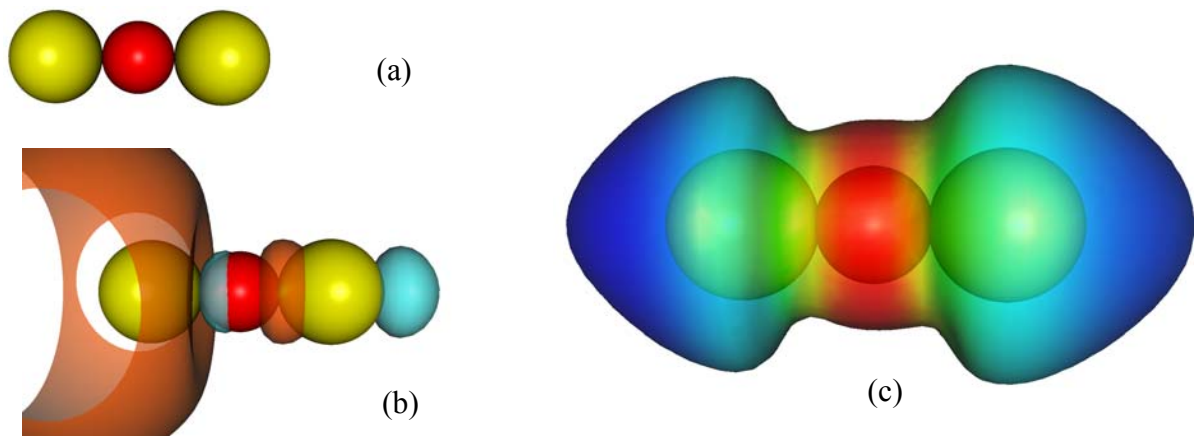

Fig. 2. (Color online) Linear pre-designed MCSCF triplet $Ni_2O$: (a) structure; (b) isosurface of the molecular electrostatic potential [MEP; values are changing from negative (red) to positive (deep blue)] corresponding to the fraction (cut) 0.01 of the maximum value (not shown) of the electron charge density distribution (CDD); (c) isosurface of the positive (red) and negative (blue) parts of the highest occupied molecular orbit (HOMO) corresponding to the isovalue 0.02. Ni and O atoms are represented by golden and red spheres, respectively. Electron charge (c) is accumulated in the vicinity of the oxygen atom (orange to yellow and to green regions). All dimensions are to scale.

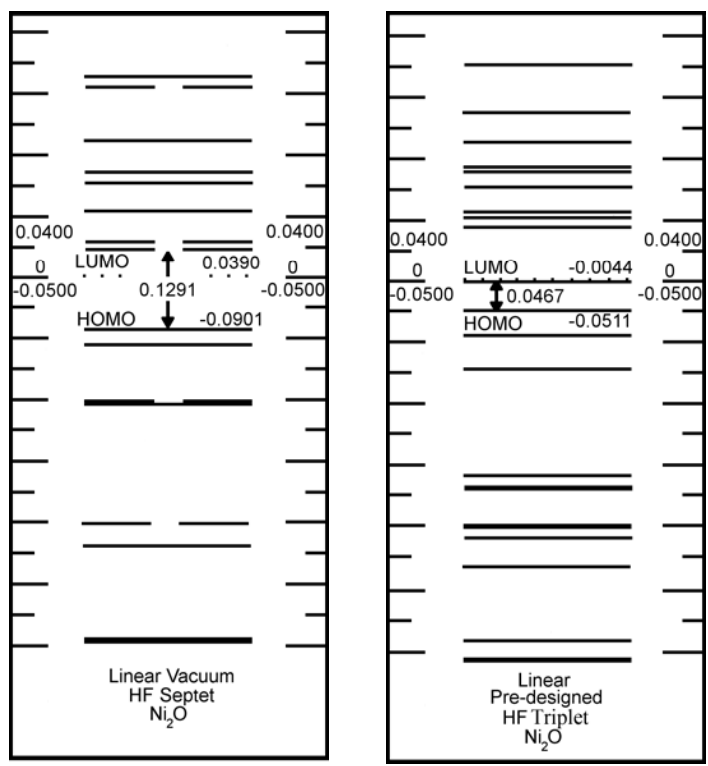

Fig. 3. HF electronic energy level structure of the vacuum (left) and pre-designed (right) linear $Ni_2O$ molecules in the HOMO-LUMO region. The scales in the unoccupied and occupied MO regions of the figures are slightly different. The electronic energies and OTEs are in Hartree units (H).

The electronic energy level structure (ELS) of these molecules is depicted in Fig. 3. Unoccupied electron energy levels in conduction bands of these molecules form numerous, closely located sub-bands, reflecting enhanced metallicity of the molecules. Valence bands of the molecules contain distinct sub-bands separated by wide gaps. The ELS structure of the linear $Ni_2O$ isomers supports a possibility of optical absorption due to $p \rightarrow d$ transitions, and is consistent with well-known experimental data (36) on optical absorption in a wide energy range for bulk NiO systems. The HF, CI and CASSCF/MCSCF data obtained in the SBKJC basis and discussed above predict a complicated spectrum of both linear $Ni_2O$ molecules in the region from about 2.5 to 4 eV, similar to that predicted for NiO molecules in the 1 to 3 eV region by the use of CI calculations with a different basis functions in Ref. 23. Interestingly enough, HF OTE of the vacuum linear $Ni_2O$ molecule is about 3.5 eV (Table I) that correlates reasonably close with CI OTE of the NiO molecule of Ref. 23, while MCSCF OTE of this molecule seem to be too large. Thus, it seems that HF calculations using SBKJC basis are as accurate as CI calculations with the basis of Ref. 23, and subsequent MCSCF calculations do not improve the HF results.

The pre-designed linear $Ni_2O$ triplet is a stable molecule that is expected to be easily synthesized experimentally in quantum confinement. While the HF absolute values of its lowest unoccupied molecular orbital (LUMO) and highest occupied molecular orbital (HOMO) energy levels are not very accurate (which is a know feature of the HF approximation), the HF OTE of this molecule (about 1.27 eV) seems reasonable in comparison with the data of Ref. 23 that show a number of local energy minima for NiO molecule with CI optical transition energies in 1 to 2 eV range. Due to non-stoichiometry, the $Ni_2O$ triplet is much more metallic than NiO molecule of Ref. 23, and thus its OTE may be smaller than that of NiO molecule. However, this assumption is not confirmed by CI/MCSCF OTE (about 4.98 eV) of the linear $Ni_2O$ triplet. The major reason for a large value of CI/MCSCF OTE of this molecule is loss of symmetry of its MOs (Fig. 3) in comparison with those of the vacuum $Ni_2O$ molecule. Indeed, spatial constraints applied to the center of mass of Ni and O atoms of the pre-designed linear $Ni_2O$ molecule destroy symmetry of its MOs as compared to those of the vacuum $Ni_2O$ molecule: the latter possesses doubly-degenerated occupied and unoccupied MOs, while the former does not (Fig. 3). The value of MCSCF OTE of the pre-designed linear $Ni_2O$ triplet also correlates very well with the fact that bulk NiO is a semiconductor with a wide gap from about 3.6 to 4.0 eV (see, for example, a discussion and references in Ref. 37). Thus, a molecule $Ni_2O$ seems likely to have OTE larger than 4.0 eV.

While the total charge of the synthesized $Ni_2O$ molecules is zero, non-stoichiometric arrangement leads to delocalization of the electron charge inside of these molecules (greenish to red regions in Figs. 1c and 2c). The Ni-containing "ends" of the molecules (greenish to blue regions in Figs. 1c and 2c) are deficient in electron charge, while "inner" regions containing O atom are slightly negative. Such electron charge and spin re-distributions are typical for non-stoichiometric molecules, and manifest violation of the standard octet rule (38, 27-31, 35).

There are at least two more spatial isomers of the studied linear $Ni_2O$ molecules: triangular vacuum and pre-designed singlets $Ni_2O$ (Table I; Figs. 4 and 5). Deep potential wells of the HF and CI/MCSCF ground states of all four $Ni_2O$ molecules correspond to the ground state energies lying within calculation error brackets from each other. The triangular vacuum singlet $Ni_2O$ has its HF ground state energy very close to that of the linear pre-designed triplet $Ni_2O$. Thus, one can conclude that the Ni-O-Ni angle of the triangular vacuum molecule may change to form a family of triangular molecules, including a degenerated one, which is the linear pre-designed molecule studied above. CI/MCSCF studies confirm, that the MCSCF ground state energies and OTEs of the triangular vacuum $Ni_2O$ molecule and those of its pre-designed linear isomer differ more than the corresponding values obtained in the HF approximation. This leads to a conclusion that both triangular vacuum and linear pre-designed molecules are stable and can be synthesized experimentally. At the same time, this confirms that the linear vacuum $Ni_2O$ molecule cannot be stable (as illustrated by the fact that it is HF septet), because the *minimum minimorum* of the total energy of the 3-atomic cluster $Ni_2O$ is realized as the triangular vacuum singlet. Moreover, the mere existence of the pre-designed triangular singlet $Ni_2O$ demonstrates, yet again, that the Ni-O bond is both stretchable and flexible, and thus may accommodate existence of various triangular conformations of $Ni_2O$ cluster, depending on confinement conditions. In conjunction with the bulk NiO system, these properties of the Ni-O bond provide for bulk NiO being a wide band gap semiconductor.

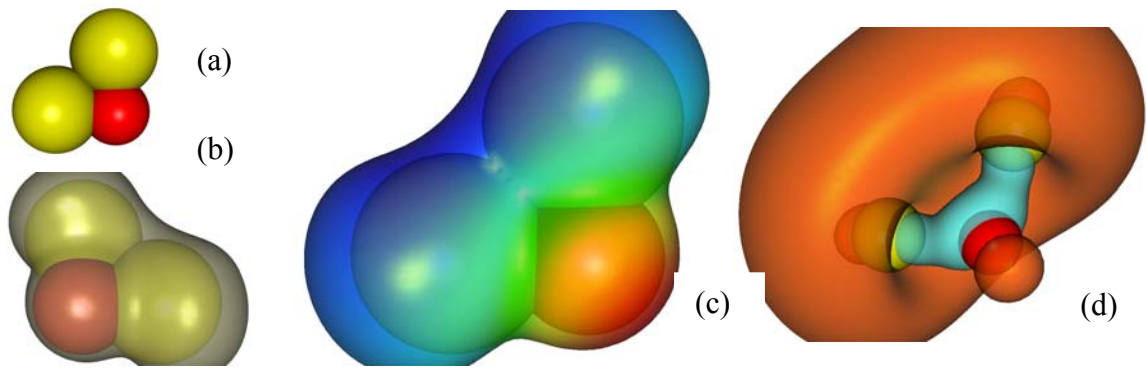

Fig. 4. (Color online) Triangular vacuum MCSCF singlet $Ni_2O$: (a) structure; (b) isosurface of the electron charge density distribution (CDD) corresponding to the fraction (cut) 0.015 of its maximum value (not shown); (c) isosurface of the molecular electrostatic potential [MEP; values are varying from negative (red) to positive (deep blue)] corresponding to the fraction (cut) 0.02 of the maximum value (not shown) of CDD; (d) isosurfaces of the positive (red) and negative (blue) parts of the highest occupied molecular orbit (HOMO) corresponding to the isovalue 0.02. Ni and O atoms are represented by golden and red spheres, respectively. Electron charge is accumulated (c) in the vicinity of the oxygen atom (orange to yellow and to green regions). Atomic dimensions in (a), (b) and (c) are to scale, and in (d) reduced. Other dimensions are to scale.

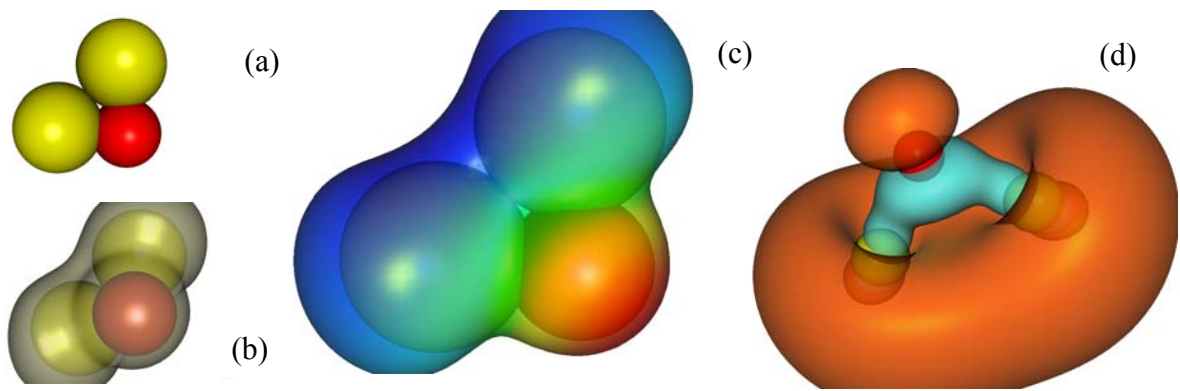

(a)

(b)

(c)

(d)

Fig. 5. (Color online) Triangular pre-designed MCSCF singlet $Ni_2O$: (a) structure; (b) isosurface of the electron charge density distribution (CDD) corresponding to the fraction (cut) 0.015 of its maximum value (not shown); (c) isosurface of the molecular electrostatic potential [MEP; values are varying from negative (red) to positive (deep blue)] corresponding to the fraction (cut) 0.02 of the maximum value (not shown) of CDD; (d) isosurfaces of the positive (red) and negative (blue) parts of the highest occupied molecular orbit (HOMO) corresponding to the isovalue 0.02. Ni and O atoms are represented by golden and red spheres, respectively. Electron charge is accumulated (c) in the vicinity of the oxygen atom (orange to yellow and to green regions). Atomic dimensions in (a), (b) and (c) are slightly enlarged, and in (d) reduced. Other dimensions in the figures are to scale.

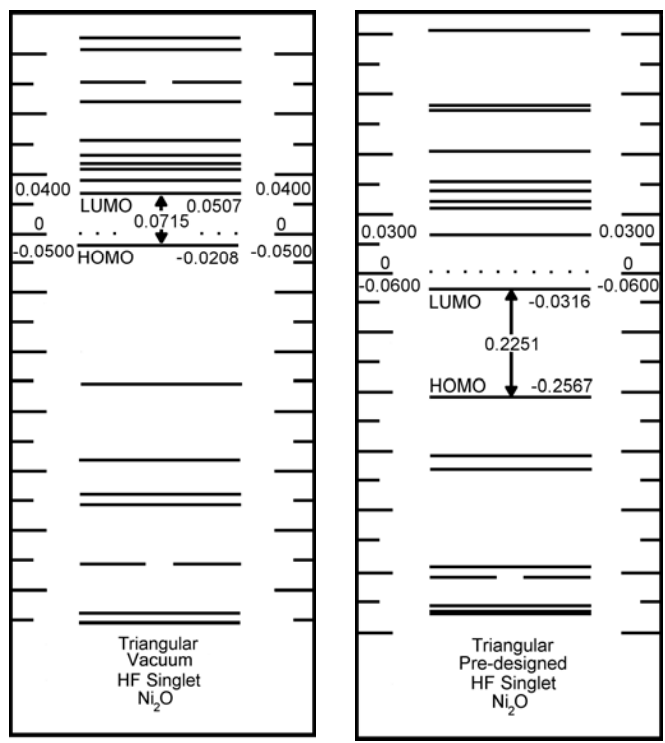

Fig. 6. HF electronic energy level structure of the vacuum (left) and pre-designed (right) triangular $Ni_2O$ molecules in the HOMO-LUMO region. The scales in the unoccupied and occupied MO regions of the figures are slightly different. The electronic energies and OTEs are in Hartree units (H).

The lengths of the Ni-O and Ni-Ni bonds (1.738 Å and 2.373 Å, respectively) in the triangular vacuum singlet (Fig. 4) are smaller than those in the predesigned one (1.900 Å and 2.480 Å, respectively, Fig. 5). Thus, it becomes obvious that the triangular shape serves to help the only O atom accommodate "extra" electron charge of Ni atoms by means of increasing the outside body angle Ni-O-Ni while still keeping Ni atoms as far from each other as possible. Accumulation of electron charge in the vicinity of O atom outside the $Ni_2O$ triangle (Fig. 4c) is further demonstrated by the significant dipole moment of this molecule reaching 2.2251 D. In the pre-designed triangular molecule separation of Ni atoms from each other and the oxygen atom is enforced by spatial constraints, and is larger than that in the vacuum triangular molecule. This leads to the corresponding increase in the ground state energy (Table I), and the dipole moment value (3.3601 D) of the pre-designed molecule.

In the linear $Ni_2O$ molecules uncompensated electron spins of $3d$ electrons of Ni atoms are parallel to each other, so the molecules are "ferromagnetic" MCSCF multiplets. In contrast, both triangular MCSCF $Ni_2O$ singlets are "antiferromagnetic," with their antiparallel electron spins localized on Ni atoms, and zero total magnetic moment. ELS patterns of the triangular molecules (Fig. 6) are very similar to those of the linear ones. They feature closely lying sub-bands of unoccupied electron energy levels that indicate enhanced metallicity, and well-developed "semiconductor"-type structure in the occupied MO regions. The symmetry of the triangular vacuum singlet breaks when the spatial constraints are applied to the center of mass of its atoms. This produces the corresponding pre-designed singlet with ELS that does not have degenerate MOs in the HOMO-LUMO region.

In correspondence with known properties of the HF approximation, it works better for the triangular vacuum singlet, which realizes the global minimum of total energy of the $Ni_2O$ cluster.

In this case HOMO and LUMO energies both have correct signs, and the OTE is close to 2 eV (Table I). In the case of the local minima of the total energy of the $Ni_2O$ cluster, which are realized by the rest of the virtually synthesized $Ni_2O$ isomers, the electron level energies in the HOMO-LUMO regions are calculated less accurately. In particular, the HOMO and LUMO energy levels seem to be too close in the case of the linear vacuum septet. For the pre-designed molecules the LUMO energy levels lie slightly below zero. Similar to the case of the linear $Ni_2O$ molecules, the MCSCF approximation does not improve HF OTE values of the triangular $Ni_2O$ molecules. Considering that metallicity of all $Ni_2O$ molecules is enhanced in comparison with that of Ni-O molecule of Ref. 23, it seems highly unlikely that the OTEs of these molecules exceed 5 eV, let alone exceeding 6 and 10 eV predicted by MCSCF calculations. This tendency correlates with findings of Ref. 23, where the use of CI (although with a less accurate basis than SBKJC one) also did not improve the OTE data obtained for NiO molecule in the HF approximation. Thus, it may happen that MCSCF approximation used with the SBKJC basis does not work well for non-stoichiometric "metallic" oxides of transition metals.

Magnetic properties of the triangular $Ni_2O$ molecules studied here by quantum many body-theoretical means are similar to those of a triangular molecule of Ref. 22 obtained using a DFT-based method, and supported by some experimental results also reported in Ref. 22. Both molecules (that of Fig. 4 and that of Ref. 22) possess antiferromagnetic arrangement of electron spins localized on Ni atoms. Unfortunately, the ground state and OTE data for the molecule of Ref. 22 are not available, so any quantitative comparison is not possible at this time.

It is also interesting to compare properties of $Ni_2O$ molecules obtained by the virtual synthesis method here with those of a vacuum NiO dimer of Ref. 25 synthesized using a similar method, the spin-unrestricted HF approximation. The length of the Ni-O bond (1.738 Å) in the triangular vacuum singlet (Fig. 4) is comparable with the length of the Ni-O bond (1.784 Å) in this dimer, while both are significantly shorter than the Ni-O bond length of 2.084 Å in the case when NiO molecule were "embedded" into a model crystal of NiO studied in Ref. 25. [The bond length of 2.084 Å is also known from earlier experiments for bulk NiO (see Ref. 39 and references therein).] This fact corroborates results obtained and discussed above for the pre-designed triangular $Ni_2O$ molecule of Fig. 5, with its Ni-O bond of 1.900 Å, and confirms that such a molecule synthesized in quantum confinement may, indeed, possess a longer Ni-O bond. [Note, that the Ni-O bond length of 1.900 Å was calculated from experimental data on covalent radii of Ni and O atoms in 12 coordinated metals (40).] Easily flexible and stretchable Ni-O bond ensures synthesis and stabilization of such molecules at various nucleation conditions. In the process of such adjustment, electron spin alignment changes dramatically from the ferromagnetic to antiferromagnetic type, and *vice versa*, while the ground state energy changes within 0.2 H, that is, within calculation error brackets.

In conclusion, it's important to note that all $Ni_2O$ molecules of this study have hybridized HOMOs composed largely of $3d$ orbits of Ni atoms, with smaller contributions due to $2p$ orbits of O atoms (Figs. 1b, 2b, 4d and 5d), similar to that of the Ni-O dimer of Walch and Goddard III (23). With an increase of the ratio of oxygen to nickel atoms in Ni - O molecules hybridization of their MOs in the HOMO-LUMO region increases, in correspondence with well-known experimental and theoretical evidence suggesting that wide gap semiconductors possess highly hybridized MOs in this region.

## 3. MOLECULES DERIVED FROM Ni₂O₂ CLUSTER

As demonstrated by studies presented in section 2, $Ni_2O$ cluster can easily adjust to nucleation conditions adopting different spatial forms to minimize its total energy, and thus producing a number of spatial isomers. In the process of such adjustment spin alignment of the cluster atoms changes dramatically (from the ferromagnetic to antiferromagnetic type), while HF OTEs change from about 1 to over 6 eV (the corresponding change in MCSCF OTE values is up to about 2 times, see Table I). This structural flexibility is further revealed in studies of this section, and is a common property of few-atomic Ni-O clusters.

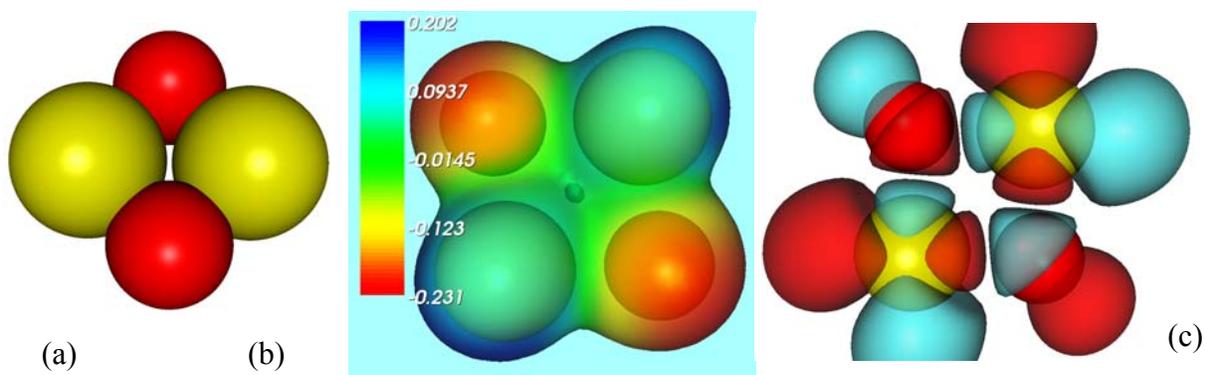

Fig. 7. (Color online) Square vacuum MCSCF pentet $Ni_2O_2$: (a) structure; (b) isosurface of the molecular electrostatic potential [MEP; values are varying from negative (red) to positive (deep blue)] corresponding to the fraction (cut) 0.05 of the maximum value (not shown) of CDD; (c) isosurfaces of the positive (red) and negative (blue) parts of the highest occupied molecular orbit (HOMO) corresponding to the isovalue 0.01. Ni and O atoms are represented by golden and red spheres, respectively. Electron charge is accumulated (c) in the vicinity of the oxygen atoms (orange to yellow and to green regions on the outer side of a portion of space occupied by the atoms). Atomic dimensions in (a) are to scale, and in (b) and (c) somewhat reduced. Other dimensions in the figures are to scale.

Yet another interesting property of small Ni-O molecules is demonstrated via comparison of properties of several isomers produced by $Ni_2O_2$ cluster (Table 1; Figs. 7 to 10). In particular, while three of four virtually synthesized $Ni_2O_2$ molecules are very stable and possess large OTEs, the predesigned molecules have their OTEs much larger than that of the corresponding vacuum molecules. This manifests a stabilizing role of environment in the process of synthesis of small Ni-O molecules.

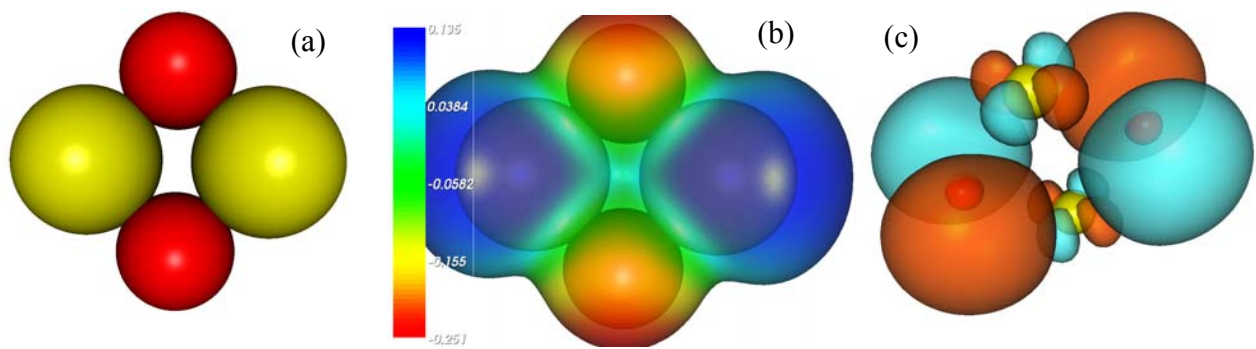

Fig. 8. (Color online) Square pre-designed MCSCF singlet $Ni_2O_2$: (a) structure; (b) isosurface of the molecular electrostatic potential [MEP; values are varying from negative (red) to positive (deep blue)] corresponding to the fraction (cut) 0.02 of the maximum value (not shown) of CDD; (c) isosurfaces of the positive (red) and negative (blue) parts of the highest occupied molecular orbit (HOMO) corresponding to the isovalue 0.01. Ni and O atoms are represented by golden and red spheres, respectively. Electron charge is accumulated (c) in the vicinity of the oxygen atoms (orange to yellow and to green regions on the outer side of a portion of space occupied by the atoms). Atomic dimensions in (a) and (b) are to scale, and in (c) reduced. Other dimensions in the figures are to

The pre-designed $Ni_2O_2$ molecule (Fig. 8) was obtained by positioning the centers of mass of Ni and O atoms in the vertices of a square with the side length equal to the sum of covalent radii of Ni and O atoms, and subsequent conditional minimization of the cluster's total energy, while the centers of mass of the oxygen and nickel atoms were kept in their initial positions. The corresponding vacuum molecule (Fig. 7) was synthesized using unconditional total energy minimization (that is, the atoms originally placed in the vertices of the square in the case of the pre-designed molecule were allowed to move, thus finally assuming their optimal positions corresponding to the global minimum of the total energy of the cluster).

Amazingly, it occurs that if one interchanges two neighboring nickel and oxygen atoms in the pre-designed $Ni_2O_2$ molecule, so the two atoms of Ni become positioned next to each other, it is still possible to minimize the total energy of the cluster so configured to obtain the modified pre-designed $Ni_2O_2$ molecule of Fig. 9 (see also Table I, No. 8). Moreover, the modified pre-designed molecule so obtained is a very stable, "antiferromagnetic" singlet with a large HF OTE of over 8 eV. The corresponding modified vacuum molecule is synthesized by application of the unconditional energy optimization to the modified pre-designed molecule. It appears to be a "ferromagnetic septet", that may be more stable than the modified pre-designed molecule, because its ground state potential well is deeper (Table I).

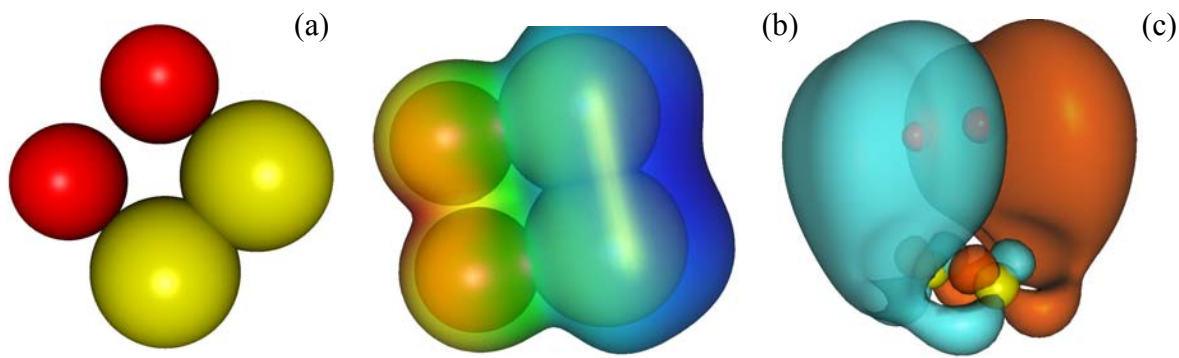

Fig. 9. (Color online) Modified pre-designed MCSCF singlet $Ni_2O_2$: (a) structure; (b) isosurface of the molecular electrostatic potential [MEP; values are varying from negative (red) to positive (deep blue)] corresponding to the fraction (cut) 0.02 of the maximum value (not shown) of CDD; (c) isosurfaces of the positive (red) and negative (blue) parts of the highest occupied molecular orbit (HOMO) corresponding to the isovalue 0.02. Ni and O atoms are represented by golden and red spheres, respectively. Electron charge is accumulated (c) between the oxygen atoms (orange to yellow and to green regions somewhat "outside" of a portion of space occupied by the cluster atoms). Atomic dimensions in (a) and (b) are to scale, and in (c) reduced. Other dimensions in the figures are to scale.

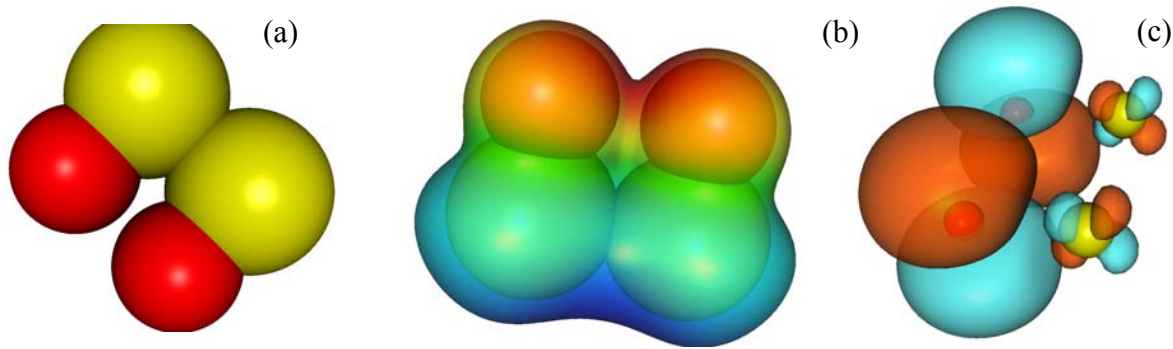

Fig. 10. (Color online) Modified vacuum MCSCF singlet $Ni_2O_2$: (a) structure; (b) isosurface of the molecular electrostatic potential [MEP; values are varying from negative (red) to positive (deep blue)] corresponding to the fraction (cut) 0.02 of the maximum value (not shown) of CDD; (c) isosurfaces of the positive (red) and negative (blue) parts of the highest occupied molecular orbit (HOMO) corresponding to the isovalue 0.02. Ni and O atoms are represented by golden and red spheres, respectively. Electron charge is accumulated (b) near the oxygen atoms (orange to yellow and to green regions somewhat "outside" of a portion of space occupied by the oxygen atoms). Atomic dimensions in (a) and (b) are to scale, and in (c) reduced. Other dimensions in the figures are to scale.

In the modified vacuum molecule (Fig. 10a) oxygen atoms are closer to the corresponding Ni atoms than those in the corresponding pre-designed molecule (Fig. 9a). In both modified molecules electron charge is accumulated in a region between the oxygen atoms somewhat outside of a space occupied by the atoms, in contrast to that of the square molecules. Thus, all modified molecules have significant dipole moments.

The ground states of all $Ni_2O_2$ molecules feature deep potential wells of about -368 H in depth, and are within the computational error brackets from each other (Table I). At the same time, HF OTEs and of these molecules are very different ranging from about 0.07 eV for the square vacuum pentet to over 8 eV for both pre-designed molecules. With an exception of the square vacuum $Ni_2O_2$ molecule, HF approximation results for these molecules seem more realistic than the corresponding MCSCF results, and the latter are consistent with MCSCF results for $Ni_2O$ molecules of sec. 2. While MCSCF ground state energies are consistent with HF ones, MCSCF OTEs seem too large to improve reasonable values of HF OTEs. In the case of the square vacuum $Ni_2O_2$ molecule MCSCF OTE value is consistent with those of the rest of the molecules, while HF one is too small, and differs over an order of magnitude from the corresponding HF results for other $Ni_2O_2$ molecules.

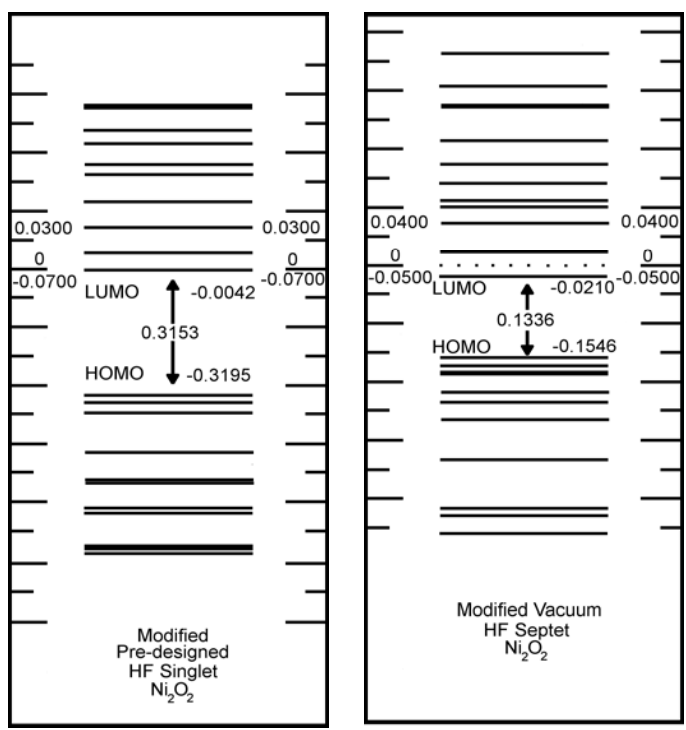

Fig. 11. HF electronic energy level structure of modified pre-designed (left) and vacuum (right) $Ni_2O_2$ molecules in the HOMO-LUMO region in the HF approximation. The scales in the unoccupied and occupied MO regions of the figures are slightly different. The electronic energies and OTEs are in Hartree units (H).

$Ni_2O_2$ molecules are stoichiometric, so the classic octet rule is expected to work for them. Indeed, HOMO of the square vacuum pentet is almost equally proportioned mixture of $3d$ and $2p$ atomic orbits of Ni and O atoms, respectively (Figs. 7c). For the modified vacuum septet (Fig. 10c), modified pre-designed singlet (Fig. 9c), and the square pre-designed singlet (Fig. 8c) contributions from $2p$ orbits of the oxygen atoms prevail. These results are consistent with the MO configuration analysis of Ref. 23 for the stoichiometric molecule NiO.

ELS of the modified $Ni_2O_2$ molecules is illustrated in Fig. 11. Location of the LUMO energy levels slightly below zero energy level is consistent with known properties of the HF approximation and confirms that it must be improved to obtain more realistic MOs and ELS. LUMO regions feature many closely lying energy levels typical for "metallic" conduction bands of small clusters, while HOMO regions are composed of bands (solid slabs) of energy levels separated by wide spaces typical for "valence" bands of small atomic clusters and molecules.

The obtained results highlight a stabilization role of quantum confinement in molecular synthesis, point out at opportunities offered by manipulations with the structure and composition of quantum confinement, and hint at possible physical mechanisms governing the exchange bias development or loss. Indeed, both pre-designed $Ni_2O_2$ molecules are stable singlets, while the corresponding vacuum molecules may be less stable pentet and septet. Moreover, both pre-designed molecules are "antiferromagnetic", while vacuum ones are "ferromagnetic" multiplets. This means that a few-atomic cluster may be stabilized into rather stable molecules at conditions where there is no "foreign" atoms within about 3 to 4 Lennard-Jones atomic diameters from the centers of mass of cluster atoms. However, such molecules may be less stable than the corresponding molecules derived from the same original atomic cluster in quantum confinement. This process may include a dramatic change of spin alignment that is crucial for the development or loss of exchange bias. As discussed in the following sections, the existence of stable Ni-O spatial isomers in quantum confinement is possible due to a stretchable and flexible Ni-O bond.

## 4. QUANTUM DOTS DERIVED FROM LARGER Ni-O CLUSTERS

Adding more nickel and oxygen atoms and exploiting other crystalline symmetries allows derivation of several larger Ni-O clusters. (Table I; Figs. 12 to 15).

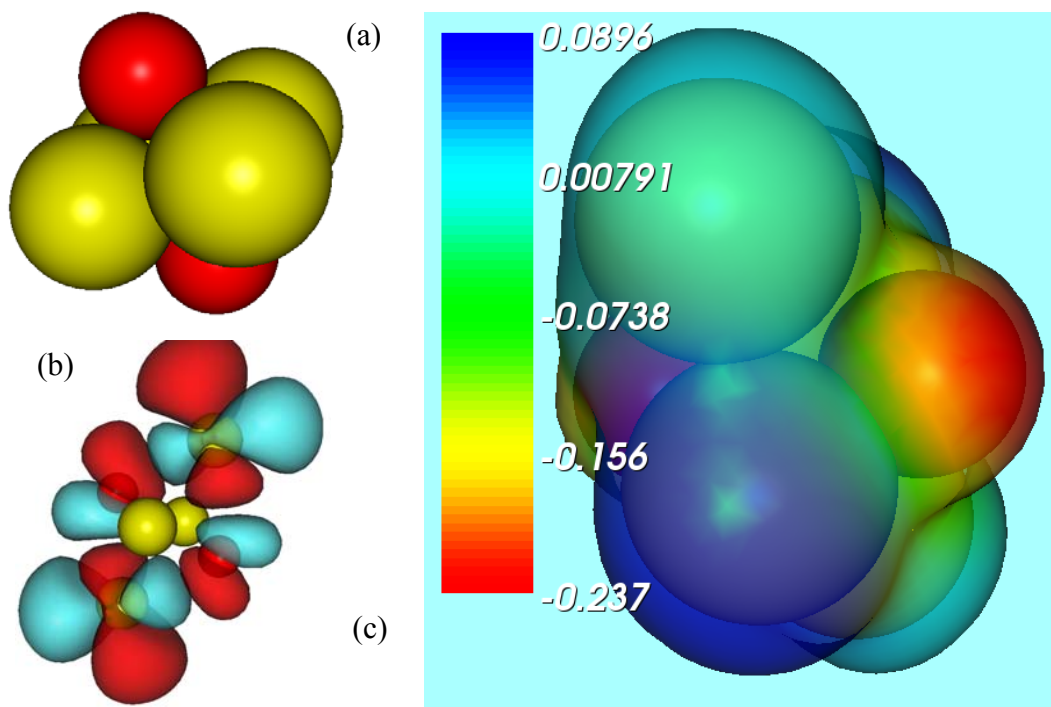

Fig. 12. (Color online) Octahedral pre-designed MCSCF septet $Ni_4O_2$: (a) structure; (b) isosurfaces of the positive (red) and negative (blue) parts of the highest occupied molecular orbit (HOMO) corresponding to the isovalue 0.005; (c) isosurface of the molecular electrostatic potential [MEP; values are varying from negative (red) to positive (deep blue)] corresponding to the fraction (cut) 0.02 of the maximum value (not shown) of CDD. Ni and O atoms are represented by golden and red spheres, respectively. Electron charge is accumulated (b) near the oxygen atoms (orange to yellow and to green regions surrounding space occupied by the oxygen atoms). Atomic dimensions in (a) are to scale, and in (b) and (c) somewhat reduced. Other dimensions in the figures are to scale.

The octahedral pre-designed septet $Ni_4O_2$ of Fig. 12 is obtained upon conditional optimization of a cluster composed of 4 Ni atoms whose centers of mass are positioned in the vertices of a square with its side length equal to two covalent radii of Ni atoms, and two oxygen atoms on the tops of two pyramids having the square as a common foundation (Fig. 12a). The pyramid side length is equal to the sum of covalent radii of Ni and O atoms. Despite of the octahedral symmetry of this molecule borrowed from bulk NiO, this molecule is a "ferromagnetic" septet that indicates a result at the edge of applicability of the HF approximation, although its ground state has a deep potential well (Table I). The HF OTE of this molecule is about 0.46 eV, and its MCSCF OTE is a bit larger (about 0.65 eV). Both values are well below the bulk NiO value, and hint again that this molecule may be unstable. The corresponding vacuum molecule also is a "ferromagnetic" septet with a parallelogram of Ni atoms in its base (as opposed to the square of Ni atoms being the base in the case the pre-designed molecule), and O atoms in the verges of the pyramids (Fig. 13) shifted with regard to each other. Its ground state potential well (Table I) is slightly less deep than that of the pre-designed molecule, but its HF OTE and MCSCF OTE values are much larger, and are similar to the CI OTE of NiO molecule of Ref. 23. Electron charge in both $Ni_4O_2$ molecules is accumulated near oxygen atoms. The vacuum $Ni_4O_2$ molecule is almost flat, with its O atoms close to the plane of Ni-based parallelogram (Fig. 13a). This tendency for vacuum molecules to become almost two-dimensional structures persists for all larger vacuum Ni-O molecules (with an exception of $Ni_4O$ and $Ni_8O_6$) of Table I, and continues for NiO quantum wires (nanopolymers) discussed in the following section.

Comparison of properties of the octahedral $Ni_4O_2$ molecules with those of the pyramidal pre-designed and vacuum molecules $Ni_4O$ helps understand whether or not the octahedral molecules are stable, and thus may be observed experimentally. The pyramidal pre-designed $Ni_4O$ pentet (Table I; Fig. 14a and 14b) studied here is simply the same structure as that of the pre-designed $Ni_4O_2$ (Fig. 12a) with one oxygen atom removed, that is, a pyramid with a square foundation built of Ni atoms. The corresponding vacuum $Ni_4O$ triplet (Table I) is obtained by unconditional optimization of the pre-designed structure. The structure of the pyramidal vacuum triplet (Fig. 14c) occurs dramatically different from that of the pre-designed molecule (Fig. 14a). The oxygen atom of this molecule moves closer to Ni atoms which arrange themselves into a spatial structure and destroy the former square foundation of the pre-designed pyramid. Thus, the final structure of this molecule is very similar to that of $CH_4$. Both $Ni_4O$ molecules have almost identical HF and MCSCF ground state energies, and their HF OTEs coincide to the third digit after the dot. One would think that they are the same molecule, but the structures and MCSCF OTE values of these molecules are dramatically different. Thus, one concludes that there exists the *minimum minimorum* of the total energy of the cluster $Ni_4O$ corresponding to the vacuum triplet, and a local minimum of the total energy (corresponding to the pre-designed pentet) located very closely to the global one. The OTEs of $Ni_4O$ molecules are rather small signifying their metallicity. Once another oxygen atom is added to these pyramids, metallicity of the emerging octahedral pre-designed and vacuum distorted $Ni_4O_2$ molecules drops significantly, and their HF OTEs increase. However, MCSCF OTE of the pre-designed octahedral $Ni_4O_2$ septet decreases comparatively to OTE of the pre-designed pyramidal $Ni_4O$ pentet. The major structural and energy characteristics revealed by this comparison of the two groups of molecules are consistent with theoretical expectations. Moreover, $Ni_4O$ molecules are HF triplet and pentet, and therefore are obtained within a range of applicability of the HF approximation. This suggests that the octahedral molecules, despite of being obtained at the verge of the HF approximation, may be stable.

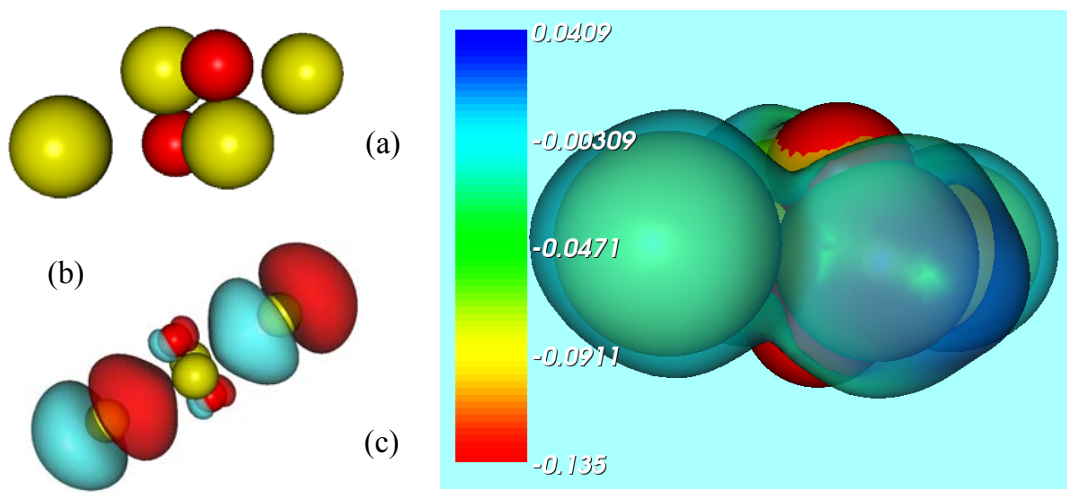

Fig. 13. (Color online) Octahedral vacuum MCSCF septet $Ni_4O_2$: (a) structure; (b) isosurfaces of the positive (red) and negative (blue) parts of the highest occupied molecular orbit (HOMO) corresponding to the isovalue 0.01; (c) isosurface of the molecular electrostatic potential [MEP; values are varying from negative (red) to positive (deep blue)] corresponding to the fraction (cut) 0.01 of the maximum value (not shown) of CDD. Ni and O atoms are represented by golden and red spheres, respectively. Electron charge is accumulated (c) near the oxygen atoms (orange to yellow and to green regions surrounding space occupied by the oxygen atoms). Atomic dimensions in (a) and (b) are reduced, and in (c) to scale. Other dimensions in the figures are to scale.

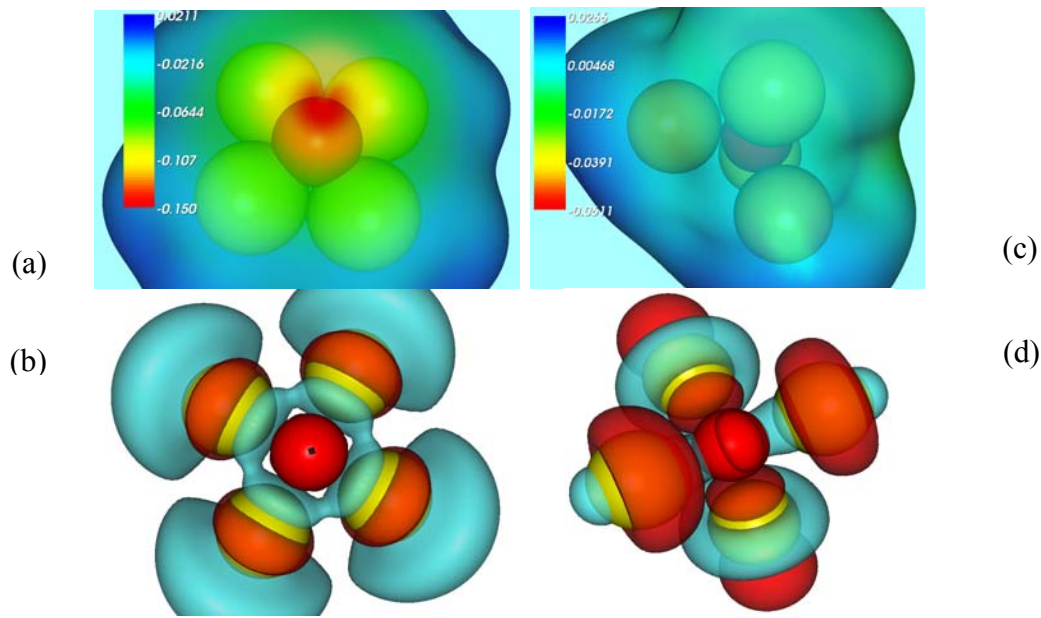

(a)

(c)

(b)

(d)

Fig. 14. (Color online) Pyramidal molecules $Ni_4O$. Isosurfaces of the molecular electrostatic potential [MEP; values are varying from negative (red) to positive (deep blue)] corresponding to the fraction (cut) 0.01 of the maximum value (not shown) of CDD for (a) pre-designed MCSCF pentet and (c) vacuum MCSCF triplet. Isosurfaces of the positive (red) and negative (blue) parts of the highest occupied molecular orbits (HOMOs) corresponding to the isovalue 0.01 for (b) pre-designed MCSCF pentet and (d) vacuum MCSCF triplet. Ni and O atoms are represented by golden and red spheres, respectively. In the pre-designed molecule electron charge is accumulated (a) near the oxygen atom (orange to yellow and to green regions surrounding space occupied by the oxygen atom). Atomic dimensions in (a) and (c) are to scale, and in (b) and (d) reduced. Other dimensions in the figures are to scale.

Many bulk semiconductors are wurtzite lattice structures containing a hexagonal prism as a symmetry element. Thus, in this work a hexagonal prism $Ni_6O_6$ cluster was optimized using conditional total energy minimization to obtain the corresponding pre-designed molecule. The pre-designed structure (Fig. 15a) is built of two "ideal" hexagons populated by alternating Ni and O atoms in their verges. One of the hexagons in this structure is turned by $\pi/3$ relatively to the rotation symmetry axis of the structure, to make O atoms of this hexagon neighboring Ni atoms of the other hexagon. Similar to the prismatic molecule of Ref. 22, the pre-designed $Ni_6O_6$ molecule is "ferromagnetic". It possesses a deep ground state energy minimum of about -1105 H, reasonable HF OTE, and a huge MCSCF OTE [further Møller-Plesset (MP-2) studies are planned to ascertain the OTE value of this molecule and that of its vacuum counterpart].

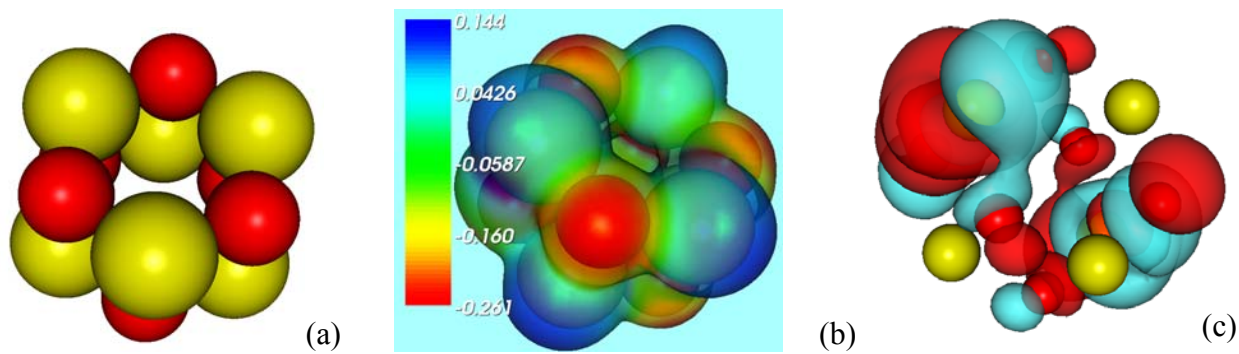

(a)                    (b)                    (c)

Fig. 15. (Color online) Prismatic pre-designed MCSCF pentet $Ni_6O_6$: (a) structure; (b) isosurface of the molecular electrostatic potential [MEP; values are varying from negative (red) to positive (deep blue)] corresponding to the fraction (cut) 0.03 of the maximum value (not shown) of CDD; (c) isosurfaces of the positive (red) and negative (blue) parts of the highest occupied molecular orbit (HOMO) corresponding to the isovalue 0.03. Ni and O atoms are represented by golden and red spheres, respectively. Electron charge is accumulated (b) near the oxygen atoms (orange to yellow and to green regions somewhat "outside" of a portion of space occupied by the oxygen atoms). Atomic dimensions in (a) and (b) are to scale, and in (c) reduced. Other dimensions in the figures are to scale.

Uncompensated electron spins of this molecule are localized on Ni atoms and produce the total magnetic moment of 4 $\mu_B$, as opposed to 12 $\mu_B$ of the DFT-derived molecule of Ref. 22. Due to stoichiometry, MEP isosurfaces corresponding to isovalues smaller than 0.01 of the pre-designed molecule (not shown in Fig. 15) reveal that the molecule is only slightly electronegative, with the electron charge evenly smoothed over its "surface". Closer to the atoms, for isovalues 0.03 and larger, MEP isosurfaces of this molecule retain a general property of small Ni-O molecules: electron charge is accumulated near oxygen atoms (Fig. 15b). A highly hybridized HOMO (Fig. 15c) of this molecule consists primarily of $3d$ atomic orbits of Ni atoms and $2p$ atomic orbits of O atoms. Significant hybridization of MOs in the HOMO-LUMO region indicates (20) that a molecule should possesses a large OTE, similar to that of insulators. Indeed, the obtained CASSCF/MCSCF OTE values for both $Ni_6O_6$ molecules are the largest OTEs of this study. Such MEP, HOMO and OTE properties manifest "semiconductor" nature of the pre-designed prismatic $Ni_6O_6$ molecule, which is the smallest "semiconductor" structure built of Ni and O atoms.

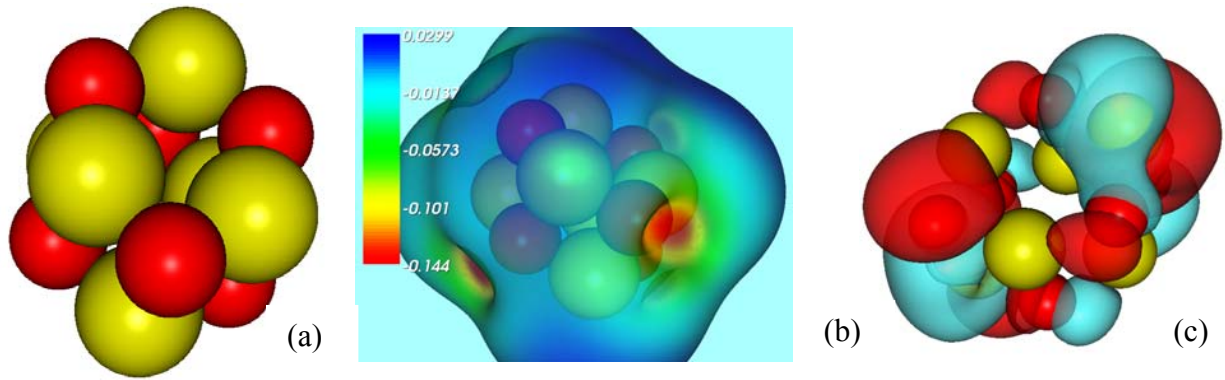

Fig. 16. (Color online) Prismatic vacuum MCSCF pentet $Ni_6O_6$: (a) structure; (b) isosurface of the molecular electrostatic potential [MEP; values are varying from negative (red) to positive (deep blue)] corresponding to the fraction (cut) 0.01 of the maximum value (not shown) of CDD; (c) isosurfaces of the positive (red) and negative (blue) parts of the highest occupied molecular orbit (HOMO) corresponding to the isovalue 0.005. Ni and O atoms are represented by golden and red spheres, respectively. Electron charge is accumulated (b) near the oxygen atoms (orange to yellow and to green regions somewhat "outside" of a portion of space occupied by the oxygen atoms). Atomic dimensions in (a) and (b) are to scale, and in (c) reduced. Other dimensions in the figures are to scale.

The prismatic vacuum $Ni_6O_6$ pentet (Table 1; Fig. 16) is obtained via unconditional total energy minimization of the pre-designed $Ni_6O_6$ cluster of Fig. 15a. The structure of this molecule is a somewhat distorted hexagonal prism where oxygen atoms moved slightly outside, and Ni atoms moved slightly inside, respectively, of their position in the ideal pre-designed hexagon. Both HF and MCSCF values of the ground state energy of this molecule are close to the corresponding values of its pre-designed counterpart, but its HF OTE of about 11 eV is much larger than that of the pre-designed molecule (in fact, it is the largest HF OTE of this study). MCSCF OTE values of both $Ni_6O_6$ molecules are comparable and both are too large, signifying that MCSCF approximation does not improve HF OTE values in this case. The electron charge of the prismatic vacuum molecule is not as evenly smoothed over the molecular "surface" as that of the pre-designed one (Fig. 16b), and is somewhat accumulated on the outside parts of regions containing oxygen atoms. Yet HOMO of this molecule is still highly hybridized, similar to that of the pre-designed molecule. Other properties of the prismatic vacuum $Ni_6O_6$ pentet (Table I, Fig. 16) are similar to those of the pre-designed one.

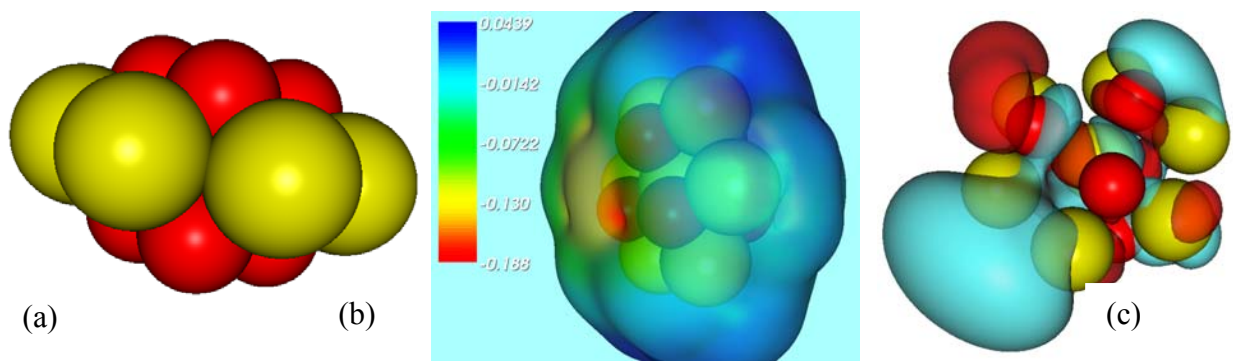

(a)                    (b)                                                      (c)

Fig. 17. (Color online) Disk-like pre-designed MCSCF pentet $Ni_7O_6$: (a) structure; (b) isosurface of the molecular electrostatic potential [MEP; values are varying from negative (red) to positive (deep blue)] corresponding to the fraction (cut) 0.01 of the maximum value (not shown) of CDD; (c) isosurfaces of the positive (red) and negative (blue) parts of the highest occupied molecular orbit (HOMO) corresponding to the isovalue 0.005. Ni and O atoms are represented by golden and red spheres, respectively. Electron charge is accumulated (b) near the oxygen atoms (orange to yellow and to green regions somewhat "outside" of a portion of space occupied by the oxygen atoms). Atomic dimensions in (a) and (b) are to scale, and in (c) reduced. Other dimensions in the figures are to scale.

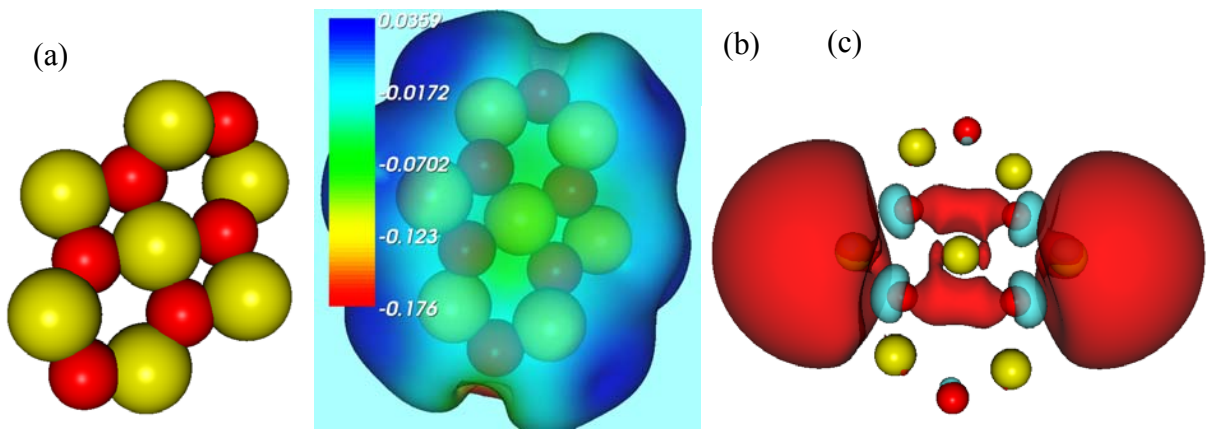

Fig. 18. (Color online) Flat, disk-like vacuum MCSCF triplet $Ni_7O_6$: (a) structure; (b) isosurface of the molecular electrostatic potential [MEP; values are varying from negative (red) to positive (deep blue)] corresponding to the fraction (cut) 0.01 of the maximum value (not shown) of CDD; (c) isosurfaces of the positive (red) and negative (blue) parts of the highest occupied molecular orbit (HOMO) corresponding to the isovalue 0.01. Ni and O atoms are represented by golden and red spheres, respectively. Electron charge is primarily accumulated (b) near two oxygen atoms at the sharper "ends" of the molecule (orange to yellow regions somewhat "outside" of a portion of space occupied by the oxygen atoms). Atomic dimensions in (a) and (b) are to scale, and in (c) reduced. Other dimensions in the figures are to scale.

The disk-like, predesigned "ferromagnetic" pentet $Ni_7O_6$ (Table I, Fig. 17) is yet another structure that realizes one of the smallest nickel oxide quantum dots (QDs). It is built of Ni atoms with their centers of mass in the verges and the center of a closed packed hexagon, and O atoms occupying three tetrahedral holes on each side of the hexagonal structure (Fig. 17a). The corresponding vacuum triplet (Table I, Fig. 18) is obtained from the pre-designed molecule using unconditional total energy minimization procedure. This vacuum molecule is almost entirely flat (fig. 18a) and somewhat twisted around its rotation symmetry axis orthogonal to the plane of this molecule shown in Fig. 18a. Its HF and MCSCF ground state energy values are slightly lower than those of the pre-designed molecule, and its small HF OTE (Table I) is still about 3 times larger than that of the pre-designed molecule. As usual in the case of Ni-O molecules, MCSCF OTEs of the disk-like pentet and triplet are sharply different from the corresponding quantities calculated in the HF approximation. The MCSCF OTE value of the vacuum triplet is about 0.4 eV, and is closer to its HF OTE value than the corresponding values in the case of the pre-designed pentet. This OTE value hints at semi-metallic nature of the vacuum $Ni_7O_6$ molecule, and may explain extraordinary chemical reactivity of experimentally observed $Ni_7O_6$ clusters (1), which is about 4 times larger than that of other Ni-O clusters.

HOMO of the pre-designed pentet (Fig. 17c) is highly hybridized, which is a fact in support of its large MCSCF OTE value, as opposed to the small OTE value obtained in the HF approximation. Both $3d$ atomic orbits of several Ni atoms, and $2p$ atomic orbits of all oxygen atoms almost equally contribute to this HOMO. In the case of the vacuum molecule, HOMO is only slightly hybridized and consists primarily of large contributions from $3d$ atomic orbits of two Ni atoms in the "ends" of the 3-atomic Ni "line" in the center of the structure, and much smaller contributions from $2p$ atomic orbits of 4 "central" oxygen atoms (Figs. 18a and 18c). The HOMO structure supports expectation of high reactivity of this molecule and explains small MCSCF and HF OTE values (Table I) specific for this molecule.

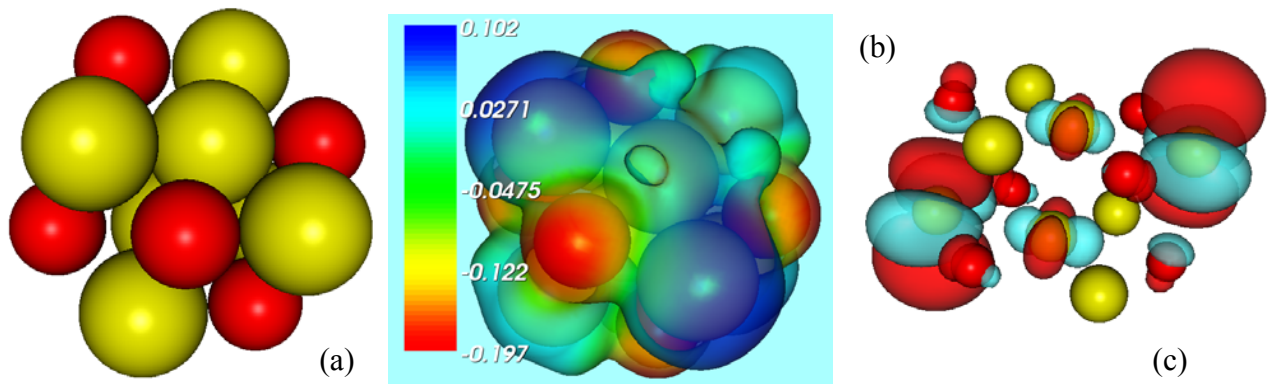

Fig. 19. (Color online) Ball-like pre-designed MCSCF triplet $Ni_8O_6$: (a) structure; (b) isosurface of the molecular electrostatic potential [MEP; values are varying from negative (red) to positive (deep blue)] corresponding to the fraction (cut) 0.02 of the maximum value (not shown) of CDD; (c) isosurfaces of the positive (red) and negative (blue) parts of the highest occupied molecular orbit (HOMO) corresponding to the isovalue 0.01. Ni and O atoms are represented by golden and red spheres, respectively. Electron charge is primarily accumulated (b) near the oxygen atoms (orange to yellow and to green regions somewhat "outside" of a portion of space occupied by the oxygen atoms). Atomic dimensions are reduced. Other dimensions in the figures are to scale.

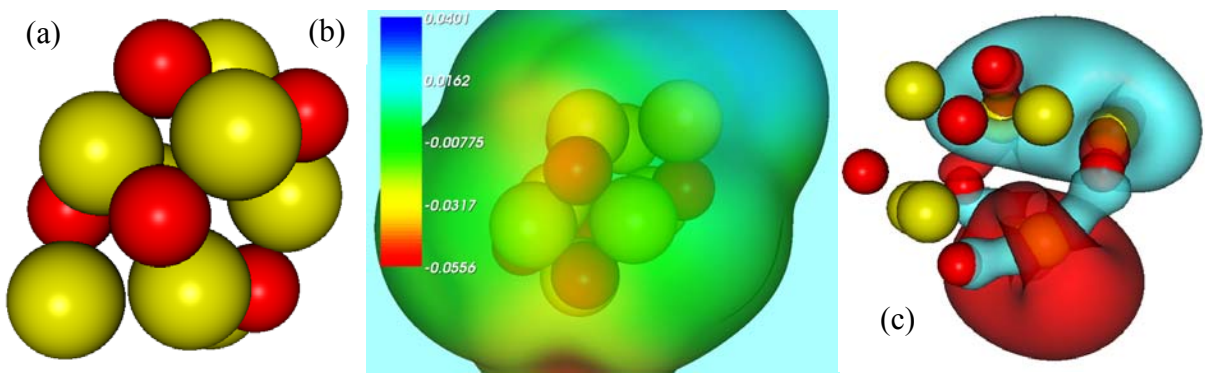

Fig. 20. (Color online) Football-like vacuum MCSCF singlet $Ni_8O_6$: (a) structure; (b) isosurface of the molecular electrostatic potential [MEP; values are varying from negative (red) to positive (deep blue)] corresponding to the fraction (cut) 0.005 of the maximum value (not shown) of CDD; (c) isosurfaces of the positive (red) and negative (blue) parts of the highest occupied molecular orbit (HOMO) corresponding to the isovalue 0.01. Ni and O atoms are represented by golden and red spheres, respectively. Electron charge is almost evenly distributed with some accumulation (b) near the oxygen atoms (orange to yellow to green regions somewhat "outside" of a portion of space occupied by the oxygen atoms). Atomic dimensions in (a) and (b) are to scale, and in (c) reduced. Other dimensions in the figures are to scale.

The largest of quantum dot-like, small Ni-O molecules studied in this study are the ball-like predesigned triplet $Ni_8O_6$ (Table I; Fig. 19) and the corresponding football-like vacuum singlet $Ni_8O_6$ (Table I; Fig. 20). They are obtained from the prismatic predesigned $Ni_6O_6$ structure (Fig. 15b) by addition of Ni atoms to the centers of each hexagon, and subsequent application of constrained and unconstrained total energy minimization procedures, respectively. The pre-designed $Ni_8O_6$ molecule recreates the smallest "cylindrical" core-shell quantum wire (QW) where a "string" of two Ni atoms, with their centers of mass on the rotation symmetry axis of the structure (Fig 20a) representing a core, is surrounded by nickel oxide shell composed of alternating Ni and O atoms with their centers of mass in the verges of two hexagons. This molecule is a "ferromagnetic" triplet with a deep ground state energy minimum (Table I) and a very small MCSCF OTE of about 0.01 eV manifesting metallicity of this QW (the HF OTE of this molecule is also small: 0.1360 eV). The corresponding "antiferromagnetic" vacuum $Ni_8O_6$ singlet is a significantly deformed version of the predesigned one (Figs. 20a and 20b), with its MCSCF OTE of about 1 eV indicating "semiconductor" nature of the vacuum QW. Its HF OTE is large: over 6.4 eV (Table I). The shape of this vacuum molecule indicates that when quantum confinement constraints are lifted, the two Ni atoms of the "string" of the ball-like pre-designed molecule (Fig.19) move apart, and the atoms in verges of the hexagons tend to "mix" around those two Ni atoms. The result is a football-like shape of the vacuum molecule that is strikingly similar to that of much larger (about 1 µm), experimentally synthesized  QWs of Ref. 42 (Fig. 21).

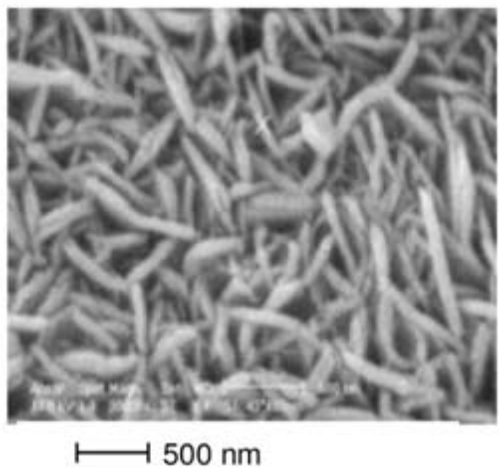

Fig. 21.  SEM image of calcinated Ni-O powder synthesized at 400ºC by a complexation-precipitation method using ammonium hydroxide as the complexation agent (from Ref. 41).

⊢————⊣ 500 nm

HOMO of the ball-like pre-designed QW $Ni_8O_6$ (Fig. 20c) is slightly hybridized and consists predominantly of contributions from $3d$  atomic orbits of 4 Ni atoms with only a small contribution from $2p$ atomic orbits of 3 oxygen atoms. This is consistent with the OTE data and indicates that the molecule is highly metallic. In contrast, HOMO of the football-like vacuum QW $Ni_8O_6$ (Fig. 20c) is a highly hybridized MO that consists of $3d$ atomic orbits of 2 Ni  atoms, and $2p$ atomic orbits of 2 oxygen atoms. This is consistent with large HF and reasonable MCSCF OTE values that signify "semiconductor" nature of this molecule.

A variety of structural solutions for small Ni-O QDs and QWs has been expected (1), and is due to Ni - O exchange interactions that produce a flexible and stretchable Ni-O bond. The RHF/ROHF ground state energies of the studied systems are close to those calculated using the CASSCF/MCSCF approximation (Table I), while the MCSCF OTEs differ significantly from HF ones, and in many cases seem to be too large. In such cases, more accurate approximations, and possibly a basis larger than SBKJC one, are necessary to ascertain OTE values of the studied molecules.

The obtained results reveal major tendencies and physical mechanisms that allow understanding experimentally observed properties of larger Ni-O systems, including large QDs, QWs and thin films. Thus, the above results indicate that as the number of Ni and O atoms grow, Ni-O molecules undergo remarkable transformation of their structure that accommodate "antiferromagnetic" and "ferromagnetic" alignment of uncompensated electron spins. About a half of the studied 8 very small molecules composed of 3 to 4 atoms (including triangular pre-designed and vacuum $Ni_2O$ molecules, pre-designed square $Ni_2O_2$ and modified pre-designed $Ni_2O_2$) are "antiferromagnetic" singlets, while the rest of the small molecules are "ferromagnetic" spin multiplets. With further growth in the number of atoms, only "ferromagnetic" spin multiplets were virtually synthesized, until the football-like vacuum $Ni_8O_6$ QW, that appears again to be a singlet. Results discussed in the following section show that father proportional increase in the numbers of Ni and O atoms in the molecules $Ni_{2+x}O_x$ with $x$ even and varying from 8 to 22 produces "antiferromagnetic" pre-designed singlets (save for only one exception) that are almost one-dimensional nanopolymer QWs up to about 6 nm in length.

In the process of nucleation and synthesis in quantum confinement or on surfaces the structural transformation described above is expected to lead to the development or loss of exchange bias, provided the confinement/surface has appropriate spin alignment of its atoms. For example, the distorted octahedral vacuum septet $Ni_4O_2$ may be physically absorbed at a Ni-O surface edge built of $Ni_{18}O_{16}$ singlet QW. The absorbed molecule interacts only with a portion of the surface that is about 3.5 Lennard-Jones atomic diameters wider in linear dimensions than the molecule's linear dimensions (the rest of the surface adds only a small contribution to this interaction). Given a broad energy plateau near the minimum of the total energy of Ni-O clusters, at some thermodynamic conditions (pressure, temperature) the spin alignment of the octahedral molecule should not change, and thus the exchange bias interaction with the substrate should develop. When thermodynamic conditions change and a deeper minimum of the total energy would be feasible, the spin alignment of the octahedral molecule may change, and it may join $Ni_{18}O_{16}$ molecule to polymerize into a pre-designed $Ni_{22}O_{18}$ singlet (QW), so that the exchange bias effects disappear.

All small vacuum molecules of Table I tend to be as flat as possible. This tendency and an ability to self-assemble in a preferential direction lead to the development of almost one-dimensional Ni-O nanopolymer QWs discussed in the following section.

## 5. Ni-O POLYMER NANOWIRES

Attempts to synthesize virtually non-stoichiometric Ni-O QWs longer than $Ni_8O_6$ of Table I and featuring hexagonal, pentagonal, square or triangular arrangement of alternating Ni and O atoms in the verges of the polygons forming QW "shells" and an axial string of Ni atoms were not successful. Instead, it was found that there exists a tendency to "self-assembly" of Ni-O molecules beginning with the pre-designed octahedral molecule $Ni_4O_2$ (No. 11 in Table I). In particular, addition of (i) three Ni atoms to each of the two rows of Ni atoms in the foundation of this molecule (Fig. 12a) and (ii) the corresponding three O atoms in the orthogonal plane to develop two "strings" of O atoms placed into octahedral holes above and below every Ni-based square as shown in Fig. 22a, with subsequent conditional minimization of the total energy of the obtained structure, lead to a pre-designed HF singlet $Ni_{10}O_8$ of Fig. 22.

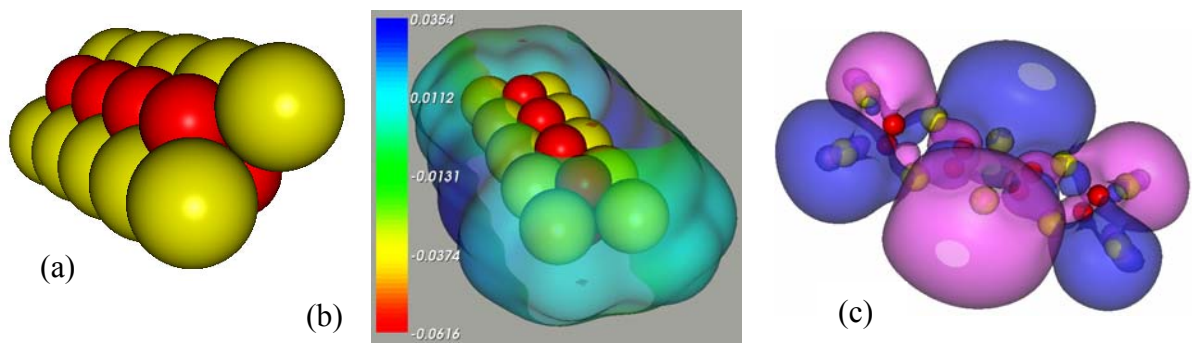

Fig. 22. (Color online) Pre-designed HF singlet $Ni_{10}O_8$: (a) structure; (b) isosurface of the molecular electrostatic potential [MEP; values are varying from negative (red) to positive (deep blue)] corresponding to the fraction (cut) 0.005 of the maximum value (not shown) of CDD; (c) isosurfaces of the positive (magenta) and negative (blue) parts of the highest occupied molecular orbit (HOMO) corresponding to the isovalue 0.005. Ni and O atoms are represented by golden and red spheres, respectively. Electron charge is almost evenly distributed around the "ends" of the molecule (b) (yellow to green regions). Atomic dimensions in (a) and (b) are to scale, and in (c) reduced. Other dimensions in the figures are to scale.

Direct evaluations using Table I data show that such "polymerization" mechanism is energetically favourable and may produce quasi one-dimensional QWs possessing remarkably deep ground state potential wells and large OTEs. Indeed, comparing HF ground state energy data for the pre-designed $Ni_7O_6$ molecule (No. 15 in Table I) with that of the pre-designed $Ni_8O_6$ triplet (No. 17 in Table I) one can find out that addition of one Ni atom to a pre-designed Ni-O structure leads to a decrease in the ground state energy by roughly 168 H. [The same result can be obtained comparing the ground state energies of any appropriate pair of molecules, such as the pre-designed molecules $Ni_6O_6$ and $Ni_8O_6$ (No. 13 and No. 17 in Table I, respectively).] Further comparison of the ground state energies of the pre-designed octahedral $Ni_4O_2$ molecule (No. 11 in Table I) and the pre-designed pyramidal pentet $Ni_4O$ (No. 10 in Table I) produces similar evaluation in the case of addition of one O atom to a pre-designed Ni-O structure: 16 H. Again, the latter evaluation can be checked by comparison of the ground state data for other appropriate pairs of molecules, such as the pre-designed molecules $Ni_2O$ and $Ni_2O_2$ (No. 4 and No. 6 in Table I, respectively), producing the same value of 16 H. Using the evaluations data discussed above, one can predict the ground state energy of the pre-designed $Ni_{10}O_8$ HF singlet of Fig. 22. Indeed, taking the ground state of the pre-designed QW $Ni_8O_6$ as the calculation basis, and subtracting from that value the change of 368 H in the total energy due to addition of two Ni and two O atoms, one can predicts that the ground state energy of the $Ni_{10}O_8$ HF singlet is about -1809 H. Amazingly, this evaluation  lies within calculation error brackets of the HF and MCSCF approximations (!) from the accurate HF value of -1808.2547340189 H (Table II).

The "polymerization" mechanism discussed above was used to pre-design and optimize a range of QWs of Table II. Beginning with QW $Ni_{10}O_8$, each subsequent QW is obtained from its predecessor by addition of two Ni and two O atoms along the length of the wire. All pre-designed QWs so polymerized  and containing from 10 to 24 Ni atoms and from 8 to 22 O atoms, respectively, are "antiferromagnetic" HF singlets with deep energy minima and OTEs about 5 eV. These QWs are 4.98 Å in width, 2.19 Å in height, and of varying length from 2.49 nm (24.9 Å) to 5.96 nm (59.6 Å). Given that QW length is over an order of magnitude larger than their width and height in all cases, the nanowires are quasi one-dimensional structures from applications' standpoint. Two examples of such Ni-O polymer nanowires are depicted in Figs. 23 and 24 below.

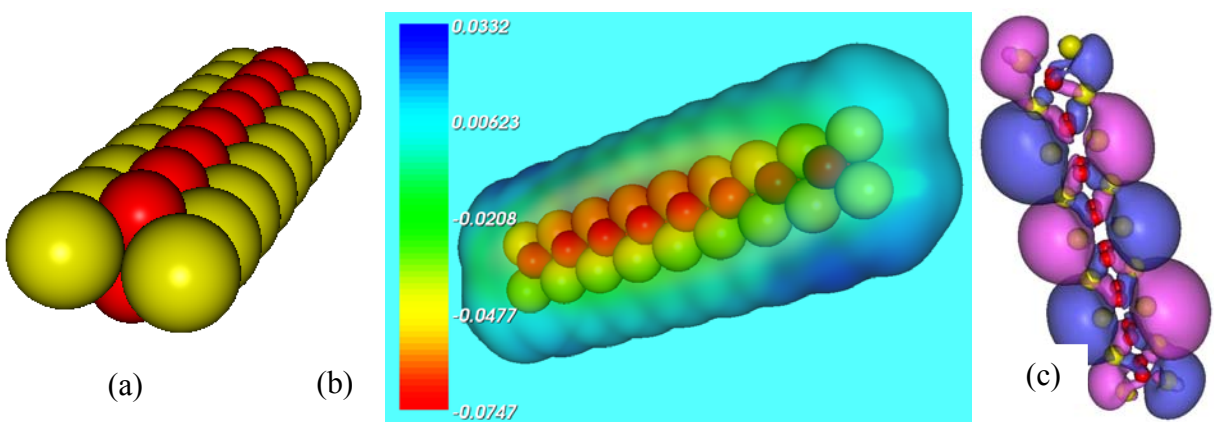

(a)      (b)      (c)

Fig. 23. (Color online) Pre-designed HF singlet $Ni_{18}O_{16}$: (a) structure; (b) isosurface of the molecular electrostatic potential [MEP; values are varying from negative (red) to positive (deep blue)] corresponding to the fraction (cut) 0.005 of the maximum value (not shown) of CDD; (c) isosurfaces of the positive (magenta) and negative (blue) parts of the highest occupied molecular orbit (HOMO) corresponding to the isovalue 0.005. Ni and O atoms are represented by golden and red spheres, respectively. Electron charge is almost evenly distributed along the "planes" of the molecule (b) (yellow to green regions). Atomic dimensions in (a) and (b) are to scale, and in (c) reduced. Other dimensions in the figures are to scale.

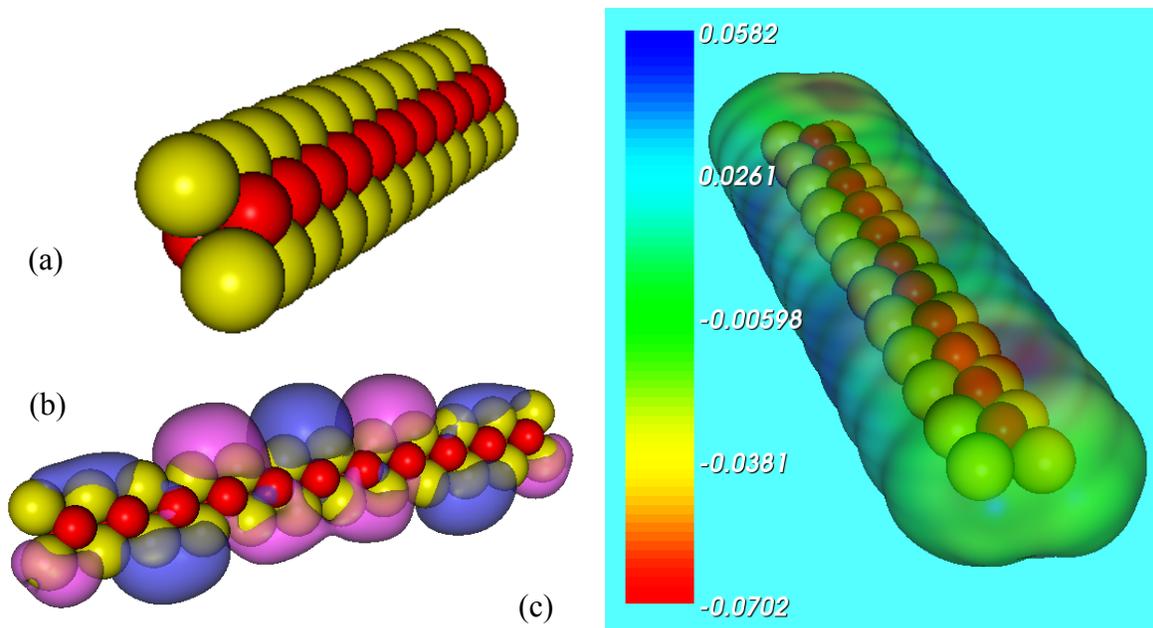

Fig. 24. (Color online) Pre-designed HF singlet $Ni_{24}O_{22}$: (a) structure; (b) isosurfaces of the positive (magenta) and negative (blue) parts of the highest occupied molecular orbit (HOMO) corresponding to the isovalue 0.005; (c) isosurface of the molecular electrostatic potential [MEP; values are varying from negative (red) to positive (deep blue)] corresponding to the fraction (cut) 0.005 of the maximum value (not shown) of CDD. Ni and O atoms are represented by golden and red spheres, respectively. Electron charge is almost evenly distributed (c) along the "planes" (green regions) of the molecule, with some accumulations in the center and near the "ends" (reddish spots and yellow to green regions). All dimensions are to scale.

All HF singlet molecules of Table II have their electronic charge smoothly distributed over the molecular "surfaces" with some accumulation over the "ends". The first two smaller molecules of Table II do not possess any other charge accumulation regions (see, for example, Fig. 22b). As the molecular size increases, electron charge begins to accumulate in the central regions of the molecular "plains" (see, for example, Fig. 23b). With further increase in the molecular size the central region of charge accumulation splits into two and then three "spots" (Fig. 24c, reddish "spots" in the middle and closer to the "ends" of the molecular "plains") on each side of the molecule located symmetrically with regard to the rotational symmetry axes of the molecules.

HOMOs of all Ni-O nanopolymer QWs are highly hybridized (Figs. 22c, 23c and 24b), with somewhat larger contributions from $3d$ atomic orbits of Ni atoms and lesser ones from $2p$ atomic orbits of O atoms. This HOMO structure agrees very well with large OTE values for all synthesized HF singlets of Table II and signifies semiconductor nature of the QWs.

A very important observation follows from the results obtained in the case of conditional and unconditional total energy minimization of $Ni_{20}O_{18}$ cluster. It occurs that there exist at least two pre-designed isomers, the HF singlet and triplet of Fig. 26 (No. 6 and No. 7) of Table II, respectively, whose ground state energy values differ by only about 0.5 H (that is, lying within calculation error brackets from each other, with the triplet's ground state energy lesser than that of the singlet). However, OTE values of the singlet and triplet differ almost by an order of magnitude: the singlet's OTE value is about 5 eV, and the triplet's one is about 0.7 eV. To clarify which of this molecules is stable and thus may be synthesized in quantum confinement experimentally, the corresponding vacuum molecule was optimized (No. 8 in Table II). The vacuum molecule occurred a HF singlet with its ground state energy value lying about 0.5 H below that of the pre-designed triplet, and its OTE value slightly larger than that of the singlet.

Comparative analysis of the obtained data for $Ni_{20}O_{18}$ QWs against those for QWs of Table II with the number of Ni atoms from 10 to 18 and O atoms from 8 to 16 leads to a conclusion that "antiferromagnetic" HF singlet states realize total energy minima for the pre-designed molecules of Table II from $Ni_{10}O_8$ to $Ni_{18}O_{16}$. Further increase in the number of Ni and O atoms produces QWs possessing almost flat total energy surfaces near the minima of the total energy corresponding to their ground states. This indicates a possibility of "polymerization" routes other than simply enlargement of the structures along one direction. Indeed, current studies of larger pre-designed QWs, $Ni_{26}O_{24}$ and $Ni_{30}O_{26}$, show that these QWs may have several isomers with their ground state energies lying within calculation error brackets from each other and that of the corresponding vacuum molecules.

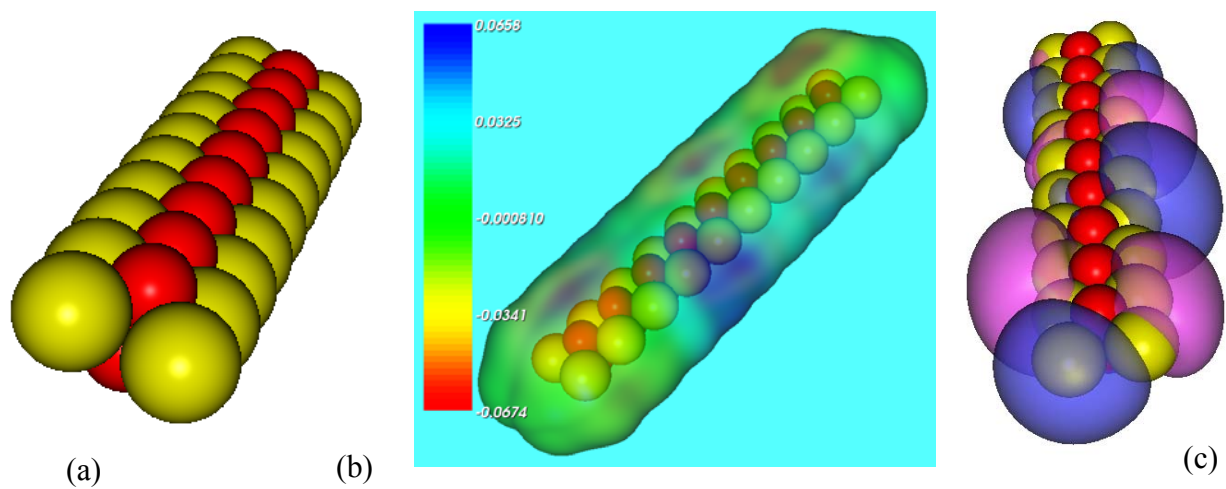

(a)      (b)      (c)

Fig. 25. (Color online) Pre-designed HF triplet $Ni_{20}O_{18}$: (a) structure; (b) isosurface of the molecular electrostatic potential [MEP; values are varying from negative (red) to positive (deep blue)] corresponding to the fraction (cut) 0.005 of the maximum value (not shown) of CDD; (c) isosurfaces of the positive (magenta) and negative (blue) parts of the highest occupied molecular orbit (HOMO) corresponding to the isovalue 0.005. Ni and O atoms are represented by golden and red spheres, respectively. Electron charge is almost evenly distributed (b) along the "planes" of the molecule (yellow to green regions), with some accumulations in the center and near the "ends" (reddish spots and yellow to green regions). All dimensions are to scale.

The above results and discussion show that further addition of Ni and O atoms to $Ni_{18}O_{16}$ molecule to obtain $Ni_{20}O_{18}$ one may or may not results in a dramatic change in uncompensated electron spin alignment from "antiferromagnetic" one in $Ni_{18}O_{16}$ molecule to "ferromagnetic" in $Ni_{20}O_{18}$ one and larger molecules. Projecting this results on a process of sorption of Ni and O atoms on Ni-O nanocluster surfaces one can conclude that in the process of cluster growth two types of larger clusters may emerge: those with antiferromagnetic and ferromagnetic spin alignment. Obviously, thermodynamic conditions must play a crucial role selecting a thermodynamicaly preferential type of spin alignment at a given range of the cluster sizes that produces a minimum of the total energy of the cluster. At the next step of materials development, such clusters may be deposited on a surface that may have one of the above types of spin alignments near a location where a cluster join that surface. In the case when an "antiferromagnetic" cluster is deposited onto a "ferromagnetic" area on a surface, or *vice versa*, exchange bias interaction between the cluster and the surface will be significant.

## 6. DISCUSSION AND CONCLUSIONS

A family of small Ni-O molecules, including the smallest oxidized core-shell QDs and QWs, have been synthesized using several progressively more accurate, many body quantum field theoretical methods. The obtained results capture the major magnetic and electronic properties of the studied molecules that point toward possible physical mechanisms behind the exchange bias development and its loss. Thus, very small triangular Ni-O molecules in confinement or on surfaces are "antiferromagnetic" singlets, and somewhat larger molecules are primarily "ferromagnetic" multiplets that tend to become triplets and singlets as the numbers of Ni and O atoms proportionally increase up to 8 Ni and 6 O atoms. The next in the family of virtually synthesized molecules, the HF singlet $Ni_{10}O_8$ (No. 1 in Table II), and all of the larger molecules of Table II, have been obtained from the pre-designed octahedral molecule $Ni_4O_2$ (No. 11 in Table I) by polymerization in a preferential direction, and occur to be one-dimensional QW singlets up $Ni_{20}O_{18}$. At this point it was found that the triplet $Ni_{20}O_{18}$ has its ground state energy slightly lesser than the corresponding singlet. While the difference in the ground state energies is within error brackets of calculations, the mere fact signifies that when the small non-stoichiometric Ni-O QWs reach a "critical" length equal to about 18 Ni atomic diameters, there may emerge some other polymerization mechanism that would lead to QW growth in two other dimensions enabling wider, two-dimensional non-stoichiometric Ni-O quantum ribbons and football-like shapes similar to those of experimental Ni-O QWs of Fig. 21 and the vacuum $Ni_8O_6$ molecule (Fig. 20), and possibly large Ni-O QDs. Nevertheless, even larger quasi one-dimensional QW singlets, such as $Ni_{24}O_{22}$ (no. 9 in Table II) and currently studied $Ni_{22}O_{20}$, $Ni_{28}O_{26}$, and $Ni_{30}O_{28}$ do exist.

The ground state energy of the vacuum singlets of Table I is lower than that of the corresponding "ferromagnetic" triplets and pentets. The opposite observation holds for predesigned singlets: their ground state energies are higher than those of the corresponding vacuum spin multiplets. Thus, projecting these results onto an experimental situation where small Ni-O molecules are deposited on a surface, one can predict that in the presence of interface, re-structuring of such predesigned "antiferromagnetic" singlets to "ferromagnetic" Ni-O molecules may be energetically favorable and easily realizable due to flexibility of the Ni-O bond. Alternatively, reconstruction of larger Ni-O multiplets to vacuum singlets (see $Ni_8O_6$ molecules, Table I) with almost the same ground state energy is also possible (see $Ni_8O_6$ molecules, Table I). Such processes may take place at antiferromagnet-ferromagnet interface between small core-shell Ni/Ni-O clusters and a confinement or surface for clusters of some critical size, leading to a decrease in the "surface" energy of the system, and also to a loss or the development of exchange bias. Examples of larger pre-designed singlets of Table II demonstrate a possible change in spin alignment of surface atoms in a vicinity of Ni and O atoms and small molecules absorbed by the surface. In particular, adding more Ni and O atoms to a surface may happen via polymerization mechanism described in the previous section. If spins of surface atoms are aligned "ferromagnetically", then such adsorption may lead to polymerization of fragments of the surface to produce large pre-designed "antiferromagnetic" singlets of Table II or similar, so that the total energy of the system decreases.

TABLE III.  Band gap width or OTE of Ni-O thin films and nanostructures: experiment and calculations.

| No. | System | Preparation method | Temperature | Band gap width or OTE, eV |
|-----|--------|--------------------|-------------|---------------------------|
| 1. | 1 μm Ni-O films, Ref. 42. | Spray pyrolysis | 350º C | 3.6 |
| 2. | 200 to 500 Å Ni-O films, Ref. 43. | Spray pyrolysis | 300º C | 3.58 to 3.4 |
| 3. | Ni-O nanowire arrays, Ref. 44. | Optical absorption | 320º C | 3.74 |
| 4. | 60 nm to 120 nm Ni-O films, Ref. 45 | Grown from solutions | 320º C | 3.25 |
| 5. | 360 to 1000 μm Ni-O films, Ref. 46. | Optical absorption | 300º C | 4.4 |
| 6. | 9 nm Ni-O nanoparticles, Ref. 47. | Thermal deposition | 450 º C | 3.56 |
| 8. | Nanostructured Ni-O, Ref. 48. | Anodic deposition | 320º C | 3.55 |
| 9. | 0.2 to 3 μm porous films of columnar 10 nm Ni-O grains, Ref. 49. | Sputtering | 320º C | 3.2 to 3.5; |
| 10. | Ni-O$_6$ clusters, Ref. 49. | Sputtering | 320º C | 4 |
| 11. | Submicron NiOx films, Ref. 50. | Spray pyrolysis | 330 to 420º C | 3.61 to 4 |
| 12. | NiO (100) and (111) surfaces, Ref. 51 | Thermal deposition | 320º C | 3.8 |
| 13. | Gradient-corrected DFT, Ref. 52 | Calculations, bulk NiO | N/A | 5.93 |

Properties of Ni-O molecules predicted here are in a general agreement with available experimental (1, 20, 22) and computational (13, 22, 26) data. It is also interesting to compare the obtained results and tendencies with experimental and computational findings for larger Ni-O systems (Table III). The majority of systems considered experimentally or evaluated by DFT methods are much larger than those considered in this work, and are polycrystalline or amorphous, as opposed to single-crystal-type systems of this work. Experimental data of Table III were obtained at temperatures of about 300º C and high, while quantum field theoretical methods used in this work allow accurate calculations at temperatures near absolute zero. Moreover, synthesis conditions in each of the cases of Table III were very different, and their details were not described in details in the corresponding publications, so it is impossible to ensure that theoretical models used in this chapter are applicable to model synthesis of the experimentally studied systems of Table III.

However, the major tendencies and values of quantities calculated in this chapter can be compared to those reavealed by experimental data of Table III. Thus, one can observe that OTEs of the molecules virtually synthesized in this work are larger than those of much larger structures evaluated from experimental data of Table III, which is a correct tendency. The theoretical OTEs of nanopolymer Ni-O QWs of Table II lie in the range from about 4.9 to 5.33 eV, and therefore are in a very good agreement with experimental data for much larger systems of Table III, for which band gaps are expected to be smaller both because of larger system size, and because the measurements were made at high temperatures. [Nanoparticles of 9 nm in linear dimensions studied in Ref. 47 (No. 6 in Table III) are still almost twice as large as the largest of systems of Table II, and they were studied at high temperatures.] The only experimental system comparable in size with those studied theoretically in this work is Ni-$O_6$ clusters of Ref. 49 synthesized at 320º C. However, the nature of this oxide molecule is very different from that of "metallic" oxides of Table I.

From experimental data of Table III it follows that synthesis conditions define the structure and properties of the synthesized systems. This observation is in a good agreement with theoretical data of Tables I and III that prove that smallest changes in synthesis conditions may significantly affect the structure of the synthesized molecules, and thus their electronic and magnetic properties. Notably, the theoretical OTEs of this work (Table II) obtained using quantum field theoretical methods are much more consistent with experimental results of Table III than the DFT-based band gap of 5.93 eV calculated for bulk NiO in Ref. 52 that largely overestimates not only the band gap of bulk NiO, but also those of small Ni-O nanostructures and molecules.

The major conclusion derived from quantum many body-theoretical results obtained and discussed in this chapter are as follows.

- Physical and chemical properties of nanoscale Ni-O QDs and QWs depend dramatically on details of the system structure and composition. In particular: (i) OTE of such systems can be changed within an order of magnitude by manipulations with the structure and composition of QDs/QWs and quantum confinement, and synthesis conditions; (ii) the octet rule does not hold because (iii) charge is re-distributed to stabilize a non-stoichiometric molecule or a molecule synthesized in quantum confinement.

- Small Ni-O molecules and polymer nanowires up to 6 nm in linear dimensions exist as "ferromagnetic" multiplets with uncompensated electron spins parallel and localized on Ni atoms, and as "antiferromagnetic" singlets with antiparallel electron spins. Depending on a particular structure, the ground state energy of "antiferromagnetic" singlets may be higher or lower than that of similar or larger "ferromagnetic" triplets, pentets and septets.

- Re-structuring of such "antiferromagnetic" singlets to "ferromagnetic" Ni-O molecules and *vice versa* can be energetically favorable and accommodated by a highly flexible/stretchable Ni-O bond. At some thermodynamic conditions, such reversible re-structuring may lead to a decrease in the "surface" energy of the antiferromagnet - ferromagnet interface, and thus to an establishment or loss of exchange bias.

- There exists a large class of "antiferromagnetic" Ni-O QWs up to 6 nm in length that are quasi one-dimensional polymers with dipole moments varying within an order of magnitude. Such nanowires can be used to create highly ordered nanoheterostructures with pre-designed dipole moment and electron spin distributions.

## ACKNOWLEDGEMENTS

Support from NSF grants DMR #0647356 and 100053, and Teragrid/NSF grants PHY09001P and DMR 100053 is gratefully acknowledged.

## REFERENCES

1.  Vann, W. D., Wagner, R. L., and Castleman, Jr., A. W., (1998). Gas-phase reactions of nickel and nickel-rich oxide cluster anions with nitric oxide. 2. The addition of nitric oxide, oxidation of nickel clusters, and the formation of nitrogen oxide anions. *J. Phys. Chem*. **A 102**, 8804-8811.

2.  Ferrari, A. M., and Pisani, C. (2008). Reactivity of non stoichiometric Ni3O4 phase supported at the Pd(100) surface: interaction with Au and other transition metal atoms. *Phys. Chem. Chem. Phys*. **10**, 1463-1470.

3.  Xu, X., Lü, X., Wang, N. Q., and Zhang, Q. E. (1995). Charge-consistency modeling of CO/NiO (100) chemisorption system. *Chem. Phys. Lett*. **235**, 541-547.

4.  Sánchez-Iglesias, A., Grzelczak, M., Rodriguez-González, B., Guardia-Girós, P., Pastoriza-Santos, I., Pérez-Juste, J., Prato, M., and Liz-Marzán, L. M. (2009). Synthesis of multifunctional composite microgels via in situ Ni growth on pNIPAM-coated Au nanoparticles. *ACS Nano* **3**, 3184-3190.

5.  Lin, H.-Y., Lee, T.-H., and Sie, C.-Y. (2008). Photocatalytic hydrogen production with nickel oxide intercalated $K_4Nb_6O_{17}$ under visible light irradiation. *International Journal of Hydrogen Energy* **33**, 4055-4063.

6.  Volkov, V. V., Wang, Z. L., and Zou, B. S. (2001). Carrier recombination in clusters of NiO. *Chem. Phys. Lett.* **337**, 117-124.

7.  Abiade, J. T., Miao, G. X., Gupta, A., Gapud, A. A., and Kumar, D. (2008). Corrigendum to "Structural and magnetic properties of self-assembled nickel nanoparticles in yttria stabilized zirconia matrix. *Thin Solid Films* **516**, 8763-8767.

8.  Miller, J. S., and Drillon M. (Eds). (2002). "Magnetism: molecules to materials III. Nanosized magnetic materials." Wiley InterScience, New York.

9.  Nakamura, R., Lee, J.-G., Mori, H., and Nakajima, H. (2008). Oxidation behaviour of Ni nanoparticles and formation process of hollow NiO. *Philosophical Magazine* **88**, 257-264.

10. Caruge, J.-M., Halpert, J. E., Bulović, V., and Bawendi, M. G. (2006). NiO as an inorganic hole-transporting layer in quantum-dot light-emitting devices. *Nano Lett.* **6**, 2991-2994.

11. Roche, B., Voisin, B., Jehl, X., Wacquez, R., Sanquer, M., Vinet, M., Deshpande, V., and Previtali, B. (2012). A tunable, dual mode field-effect or single electron transistor. *arXiv*: 1201.3760v1 [cond-mat.mes-hall].

12. Boris, A. V., Matiks, Y., Benckiser, E., Frano, A., Popovich, P., Hinkov, V., Wochner, P., Castro-Colin, M., Detemple, E., Malik, V. K., Bernhard, C., Prokscha, T., Suter, A.,

Salman, Z., Morenzoni, E., Cristiani, G., Habermeier, H.-U., and Keimer, B. (2011). Dimensionality control of electronic phase transition in nickel-oxide superlattices. *Science* **332**, 937-940.

13.  Irwin, M. D., Buchholz, D. B., Hains, A. W., Chang, R. P. H., and Marks, T. J. (2008). p-Type semiconducting nickel oxide as an efficiency-enhancing anode interfacial layer in polymer bulk-heterojunction solar cells. *Proc. Nat Acad. Sci.* **105**, 2783-2787.

14.  Nachman, M., Cojocaru, L. N., and Ribco, L. V. (2006). Electrical properties of non-stoichiometric nickel oxide. *Phys. Stat. Solidi (b)* **8**, 773-783.

15.  Allouti, F., Manceron, L., and Alikhani, M. E. (2009). On the performance of the hybrid TPSS meta-GGA functional to study the singlet open-shell structures: A combined theoretical and experimental investigation of $Ni_2O_2$ molecule. *J. Mol. Structure. Theochem* **903**, 4-10.

16.  Nogués, J., Sort, J., Langlais, V., Skumryev, V., Surinach, S., Muñoz, J. S., and Baró, M. D. (2005). Exchange bias in nanostructures. *Phys. Reports* **422**, 65-117.

17.  Chakhalian, J., Millis, A. J., and Rondinelli, J. (2012). Whither the oxide interface. *Nature Materials* **11**, 92-94.

18.  Hwang, H. Y., Iwasa, Y., Kawasaki, M., Keimer, B., Nagaosa, N., and Tokura, Y. (2012). Emergent phenomena at oxide interfaces. *Nature Materials* **11**, 103-113.

19.  (a) Koppens, F. H. L., Buizert, C., Tielrooij, K. J., Vink, I. T., Nowack, K. C., Meunier, T., Kouwenhoven, L. P., and Vandersypen, L. M. K. (2006). Driven coherent oscillation of a single electron spin in a quantum dot. *Nature* **442**, 766-771.

(b) Loth, S., Baumann, S., Lutz, C. P., Eigler, D. M., Heinrich, A. J. (2012). Bistability in atomic-scale antiferromagnets. *Science*, **335**, 196-199.

20. Kodama, R. H., Makhlouf, S. A., and Berkowitz, A. E. (1997). Finite size effects in antiferromagnetic NiO nanoparticles. *Phys. Rev. Lett*. **79**, 1393-1396.

21. Dobrynin, A. N., Temst, K., Lievens, K., Margueritat, J., Gonzalo, J., Afonso, C. N., Piscopiello, E., and Van Tandeloo, G. (2007). Observation of Co/CoO nanoparticles below the critical size for exchange bias. *J. Appl. Phys*. **101**, 113913 (7).

22. Yi, J. B., Ding, J., Feng, Y. P., Peng, G. W., Chow, G. M., Kawazoe, Y., Liu, B. H., Yin, J. H., and Thongmee, S. (2007). Size-dependent magnetism and spin-glass behavior of amorphous NiO bulk, clusters and nanocrystals: experiment and first-principle calculations. *Phys. Rev*. **B 76**, 224402 (5) (2007).

23. Walch, S. P., and Goddard III, W. A. (1978). Electronic states of NiO molecule. *J. Amer. Chem. Soc*. **100**, 1338-1348.

24. Fujimori, A., and Minami, F. (1984). Valence-band photoemission and optical absorption in nickel compounds. *Phys. Rev*. **B 30**, 957-971.

25. Xu, X., Nakatsuji, H., Ehara, M., Lu, X., Wang, N. Q., and Zhang, Q. E. (1998). Cluster modeling of metal oxides: the influence of the surrounding point charges on the embedded cluster. *Chem. Phys. Lett*. **292**, 282-288.

26. Kadossov, E. B., Kaskell, K. J., and Langell, M. A. (2007). Effect of surrounding point charges on the density functional calculations of $Ni_xO_x$ clusters ($x$=4-12). *J. Comput. Chem*. **28**, 1240-1251.

27. Pozhar, L. A. (2010). Small InAsN and InN clusters: electronic properties and nitrogen stability belt. *Eur. Phys. J*. **D 57**, 343-354.

28. Pozhar, L. A., Yeates, A. T., Szmulowicz, F., and Mitchel, W. C. (2005). Small atomic clusters as prototypes for sub-nanoscale heterostructure units with pre-designed charge transport properties. *Eur. Phys. Lett.* **71**, 380-386.

29. Pozhar, L. A., Yeates, A. T., Szmulowicz, F., and Mitchel, W. C. (2006). Virtual synthesis of artificial molecules of In, Ga and As with pre-designed electronic properties using a self-consistent field method. *Phys. Rev*. B **74**, 085306 (11).

30. Pozhar, L. A., and Mitchel, W. C. (2009). Virtual synthesis of electronic nanomaterials: fundamentals and prospects. *In*: "Toward Functional Nanomaterials, Lecture Notes in Nanoscale Science and Technology" (Z. Wang, Ed.), Vol. **5**, pp. 423-474. Springer, New York.

31. Pozhar, L. A., and Mavromichalis, C. (2010). Spin alignment in, and electronic and magnetic properties of small Co-O molecules. *J. Appl. Phys.* **107**, 09D708 (3).

32. Schmidt, M. W., Baldridge, K. K., Boatz, J.A., Elbert, S.T., Gordon, M.S., Jensen, J.H., Koseki, S., Matsunaga, N., Nguyen, K.A., Su, S., Windus, T.L., Dupuis, M., Montgomery, J.A. (1993). General Atomic and Molecular Electronic Structure System. *J. Comput. Chem*. **14**, 1347-1363. http:// www.msg.ameslab.gov/GAMESS

33. Gordon, M. S., and Schmidt, M. W. (2005). Advances in electronic structure theory: GAMESS a decade later. *In* "Theory and Applications of Computational Chemistry: the First Forty Years" (C. E. Dykstra, G. Frenking, K. S. Kim, G. E. Scuseria, Eds.), pp. 1167-1189. Elsevier, Amsterdam. http:// www.msg.ameslab.gov/GAMESS

34. Stevens, W. J., Krauss, M., Basch, H., and Jasien, P. (1992). Relativistic compact effective potentials and efficient, shared-exponent basis sets for the third-, fourth-, and fifth-row atoms. *Can. J. Chem*. **70**, 612-630.

35. Pozhar, L. A., and Mitchel, W. C. (2007). Collectivization of electronic spin distributions and magneto-electronic properties of small atomic clusters of Ga and In with As, V and Mn *IEEE Trans. Magn.* **43**, 3037-3039.

36. Powell, R. J., and Spicer, W. E. (1970). Optical properties of NiO and CoO. *Phys. Rev.* B **2**, 2182-2193.

37. Sasi, B., and Gopchandran, K. G. (2007). Nanostructured mesoporous nickel oxide thin films. *Nanotechnology* **18**, 115613.

38. Boldyrev, A. I., Wang, L.-S. (2001). Beyond classical stoichiometry: experiment and theory. *J. Phys. Chem.* A **105**, 10759-10775.

39. Wyckoff, R. W. G. (1963). "Crystal Structures". Wiley, New York.

40. Kittel., C. (2005). "Introduction to Solid State Physics", 8th ed. Wiley, New York, p.71.

41. Kashani Moltagh, M. M., Youzbashi, A. A., and Sabaghadeh, L. (2011). Synthesis and characterization of nickel hydroxide/oxide nanoparticles by the complexation-precipitation method. *Int. J. Phys. Sci.*, **6**, 1471-1476.

42. Mahmoud, A. A., Akl, A. A., Kamal, H., and Abdel-Hady, K. (2002). Opto-structural, electrical and electrochromic properties of crystalline nickel oxide thin films prepared by a spray pyrolysis. *Physica* B, **311**, 366-375.

43. Patil, P. S., and Kadam, L. D. (2002). Preparation and characterization of spray-pyrolyzed nickel oxide (NiO) thin films. *Appl. Surf. Sci.* **199**, 211-221.

44. Lin, Y., Xie, T., Cheng, B., Geng, B., and Zhang, L. (2003). Ordered nickel oxide nanowire arrays and their optical absorption properties. *Chem. Phys. Lett.* **380**, 521-525.

45. Varkey, A. J., and Fort, A. F. (1993). Solution growth technique for deposition of nickel oxide thin films. *Thin Solid Films* **235**, 47-50.

46. Doyle, W. P., and Lonergau, G. A. (1958). Optical absorption in the divalent oxides of cobalt and nickel. *Discuss. Faraday Soc*. **26**, 27-33.

47. Wang, X., Song, J., Gao, L., Jin, J., Zheng, H., and Zhang, Z. (2005). Optical and electrochemical properties of NiO via thermal decomposition of nickel oxalate nanofibers (NiC$_2$O$_4$). *Nanotechnology* **16**, 37.

48. Boschloo, G. and Hagfeldt, A. (2001). Spectroelectrochemistry of nanostructured NiO. *J. Phys. Chem*. B, **105**, 3039-3034.

49. Wruck, D. A., and Rubin, M. (1993). Structure and electronic properties of electrochromic NiO films. *J. Electrochem. Soc*. **140**, 1096-1104.

50. Desai, J. D., Min, S. K., Jung, K. D., and Joo, O.-S. (2006). Spray pyrolytic synthesis of large area NiO$_x$ films from aqueous nickel acetate solutions. *Appl. Surface Sci*. **253**, 1781-1786.

51. Cappus, D., Xu, C., Ehrlich, D., Dillmann, B., Ventrice, C. A., Jr., Al Shamery, K., Kuhlenbeck, H., Freud, H.-J. (1993). Hydroxyl groups on oxide surfaces: NiO (100), NiO (111) and Cr$_2$O$_3$ (111). *Chem. Phys*. **177**, 533-546.

52. Bredow, T., and Gerson, A. R. (2000). Effect of exchange and correlation on bulk properties of MgO, NiO and CoO. *Phys. Rev. B*, **61**, 5194-5201.